\documentclass[preprint,floats,aps,showpacs,epsf]{revtex4}
\usepackage{amssymb}
\usepackage{latexsym}
\usepackage{amsmath}
\usepackage{epsfig}
\usepackage{color}
\usepackage{graphicx}
\usepackage{hyperref}
\numberwithin{equation}{section}
\numberwithin{figure}{section}
\numberwithin{table}{section}

\newcommand{\ta}{\theta}
\newcommand{\be}{\begin{equation}}
\newcommand{\ee}{\end{equation}}
\newcommand{\bea}{\begin{eqnarray}}
\newcommand{\eea}{\end{eqnarray}}

\newcommand{\p}{\partial}
\newcommand{\s}{\sigma}

\newcommand{\la}{\langle}
\newcommand{\ra}{\rangle}
\newcommand{\rd}{\mbox{d}}
\newcommand{\ri}{i}
\newcommand{\re}{\mbox{e}}

\begin{document}
\title{Particle Formation and Ordering in Strongly Correlated Fermionic Systems: 
Solving a Model of Quantum Chromodynamics }

\author{ P. Azaria$^{1}$, R. M. Konik$^{2}$, P. Lecheminant$^{3}$,  T. 
P\'almai$^4$, G. Tak\'acs$^{4,5}$, and A.  M. Tsvelik$^{2}$}
\affiliation{ $^{1}$  Laboratoire de Physique Th\'eorique de Matiere 
Condens\'ee, Universit\'e Pierre et Marie Curie, CNRS, 4 Place Jussieu, 75005 
Paris, France\\
$^{2}$ Condensed Matter Physics and Materials Science Division, Brookhaven 
National Laboratory, Upton, NY 11973-5000, USA\\
$^{3}$ Laboratoire de Physique Th\'eorique et
 Mod\'elisation, CNRS UMR 8089, Universit\'e de Cergy-Pontoise, Site de 
Saint-Martin,
F-95300 Cergy-Pontoise Cedex, France,\\
$^4$ MTA-BME ``Momentum Statistical Field Theory Research Group, H-1111 
Budapest, Hungary, Budafoki \'ut 8,\\
$^5$ Department of Theoretical Physics, Institute of Physics, Budapest 
University of Technology and Economics, H-1111, Budapest, Hungary, Budafoki \'ut 
8.}
\date{\today}

\begin{abstract}
 In this paper we study  a (1+1)-dimensional version of the famous 
Nambu-Jona-Lasinio model of quantum chromodynamics (QCD2) both at zero and 
finite baryon density. We use non-perturbative techniques (non-abelian 
bosonization and the truncated conformal spectrum approach (TCSA)).  When the baryon 
chemical potential, $\mu$, is zero, we 
describe a formation of fermion three-quark (nucleons and
 $\Delta$-baryons) and boson (two-quark mesons, six-quark deuterons) bound 
states.  We also study at $\mu=0$ a formation of a topologically nontrivial phase. When the 
chemical potential exceeds the critical value and a finite baryon density 
appears, the model  has a rich phase diagram which includes phases with density 
wave and superfluid quasi-long-range (QLR) order as well as a phase of a baryon 
Tomonaga-Luttinger liquid (strange metal). The QLR order results in either a 
condensation of scalar mesons (the density wave) or six-quark bound states 
(deuterons).

\end{abstract}

\maketitle


\section{ Introduction}  

The famous Nambu-Jona-Lasinio (NJL)
model \cite{NJL} of particle physics is frequently used  as a simplified model 
of Quantum Chromodynamics (QCD) where 
the Yang-Mills force is replaced by a point-like four fermion current-current 
interaction.
The replacement does not affect the low energy sector of the theory. 

 The  (1+1)-dimensional version of the NJL model along with its prototype QCD2 
has been a subject of intense interest (see, for instance Refs. 
\cite{reviewF,bookqcd} for reviews and Ref. \cite{QCD2} for more recent work) since here one can  
obtain non-perturbative results solving  the problem of many-body bound states 
and obtaining a reliable description of the excitation spectrum. However, few 
results are available for the formation of baryons in the (1+1)-D NJL  model 
with a
bare quark mass which breaks the  U(1) chiral symmetry of the model. 
This problem has been mostly investigated in the past by variational approaches 
\cite{salcedo} and in the large-$N$ limit  for the model with an SU($N$) 
symmetry \cite{thies}.

\begin{figure}[!ht]
\centering
\includegraphics[width=0.6\columnwidth,clip]{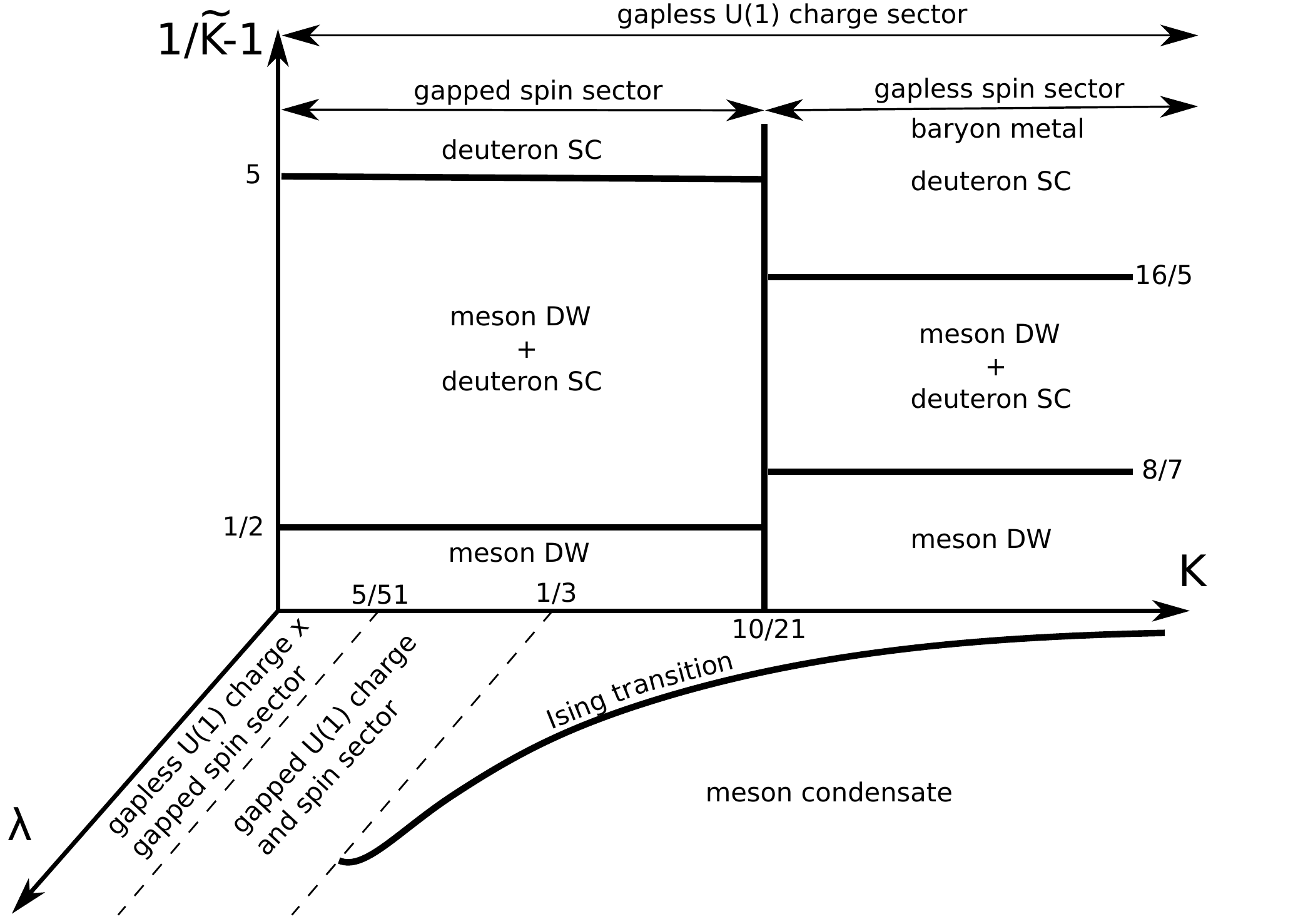}
\caption{The phase diagram of the NJL model as a function of the parameters $K, 
\tilde K$ (Luttinger parameters at zero and finite chemical potential 
respectively) and $\lambda$ (the coupling constant of the instanton term).  Here $\tilde K^{-1}-1$
can be thought of as a proxy for the density of baryons in the Fermi sea.  The $\tilde K=1$, $\lambda=0$
axis is marked by three transitions.  For $K<5/51$, the theory is gapped in the (iso)-spin sector
but gapless in the charge sector (treating the nominally irrelevant $m^*$ perturbation at second order in perturbation
theory).
In contrast for $5/51<K$, the relevant $m^*$ perturbation leads to the theory being gapped in both 
of these sectors.  For $K>1/3$ the instanton term becomes relevant.  
At finite $\lambda>0$ and $K>1/3$, the $K-\lambda$ plane is divided into two by an Ising-like transition.
The ordered side of the transition is characterized by a meson condensate.
The $\tilde K^{-1}-K$ plane corresponds to the NJL model at finite baryon density induced by a chemical
potential coupling to the U(1) charge.
For $K<10/21$ the charge sector is gapless, but the spin sector is again gapped out
by a perturbation generated at second order (note however that the value of $K$ at which this happens differs
from the zero baryon density phase).  For $K>10/21$ we have a genuine baryon metal where
both the spin and charge sectors are gapless.  At finite baryon density the systems has different
dominant quasi-long range orders, either deuteronic superconductivity (deuteron SC) or a 
meson density wave (meson DW).
In certain regions of the finite density phase diagram, the instability to both of these orders is present
(see text).}
\label{phasediag}
\end{figure}

In this paper, we study the non-perturbative spectrum and the zero-temperature 
phase diagram of the massive (1+1)-D NJL model with two flavors (SU(3)$_{\rm 
color}\times$SU(2)$_{\rm isospin}\times$U(1) symmetry). 
Our approach is based on non-abelian bosonization, conformal field theory (CFT) 
techniques \cite{bookboso,dms}, and non-perturbative numerical calculations 
based on the truncated conformal spectrum approach  (TCSA) 
\cite{truncated,konikadamov,konik}. Although non-abelian bosonization was  
applied to QCD2 before \cite{gepner,affleckqcd,bookqcd},  its combination with 
TCSA is new and this allows us to obtain a comprehensive description of the problem 
beyond the semiclassical and large color number approximations. The new results 
include  bosonized expressions for the baryon operators, nonperturbative results 
for the excitation spectrum in the zero chemical potential  regime, and the 
phase diagram (see Fig. \ref{phasediag}).

At zero baryon density (zero chemical potential), the model has two gapped phases 
separated by an Ising quantum critical point corresponding to the spontaneous 
breaking of a Z$_2$ symmetry. The order parameter is the $\gamma_5$ quark mass.
In the language of condensed matter physics this corresponds to a spontaneous dimerization  
of an insulator \cite{nersesyan}.  This breaking emerges when the t'Hooft term \cite{t'hooft} has positive coupling 
and is sufficiently strong. In the disordered phase the low energy spectrum 
consists of isospin $I=1/2$ and $I=3/2$ baryons (three-quark bound states), 
$I=0,1$ six-quark bound  states (deuterons and dibaryons), and at least eight 
mesons (quark-antiquark bound states).  The TCSA results for their masses are 
given in Table \ref{tabmasses}.

When the chemical potential exceeds some critical value so that the density of 
baryons becomes finite, the phase diagram becomes more complicated. It 
includes several phases, each with a different quasi long range (QLR) order. Depending 
on the strength of the forward scattering and the hadron density, one may have a 
phase where the SU(2) isospin sector is gapped and hence the QLR does not 
originate from a Fermi surface instability, but emerges as a Bose condensation of 
the scalar mesons (a $2k_F$ density wave (DW)) or the deuterons (superfluidity). The latter instability becomes dominant when the density 
exceeds some critical limit. In (1+1)-D there is no color superconductivity due 
to pairing of quarks contrary to what has been found in higher dimensions 
\cite{color}, but only the aforementioned superfluid phase of colorless six-quark bound 
states (the deuterons) as it was found in one-dimensional SU($N$) cold fermions models 
\cite{phle}.
There is also a critical phase  (a baryon strange metal) of gapless baryons 
forming a Tomonaga-Luttinger liquid with  gapped  color degrees of freedom.

The main body of this paper is divided into  four sections.
In Section II we discuss the NJL model and the 
sigma model which emerges as its low energy equivalent. Here we also discuss the 
bosonized form of  fermionic operators responsible for creation of mesons, 
baryons, and baryon bound states. In Section III we provide a simplified analysis 
of the excitation spectrum by means of a semiclassical approximation. In Section V,
which contains most of our new results, we analyze the spectrum using TCSA.  In 
Section IV we discuss the case of finite particle density. This section also contains 
a description of the model phase diagram. 

\section{The model} 

The NJL model describes fermionic quarks with a bare mass $m$ interacting via a 
current-current interaction. In (1+1)-dimensions, a Dirac spinor has two components 
corresponding to right- and left moving quarks. The Hamiltonian density of the left and right movers is
\begin{equation}
 {\cal H}=  \ri(-R^\dagger_{j\s}\p_x R_{j\s} + L^\dagger_{j\s}\p_x L_{j\s}) + 
m(L^\dagger_{j\s}R_{j\s} + {\rm H.c.}) +
  gJ^A\bar J^A + g_f \mathcal{J}\bar{\mathcal{J}},
\label{model}
\end{equation}
where $R_{j\s}, L_{j\s}$ are annihilation operators of the right- and the left 
moving quarks,  $j=1,2,3$ are color indices while $\s = \uparrow, \downarrow$ are flavor indices
corresponding to up and down quarks (we neglect all others in this treatment). 
The speed of light is set to one and a summation over repeated 
indices is implied in the following. 

We begin the treatment of this Hamiltonian through recourse to non-abelian bosonization (for additional
details see Appendix A).  To
this end we identify the SU(3)$_2$ Kac-Moody (KM)  
currents of right and left chirality: 
\[
J^A = :R^\dagger_{j\s} T^A_{jk} R_{k\s}:, \bar J^A = :L^\dagger_{j\s} T^A_{jk} L_{k\s}:,
\]
where $T^A_{jk}$ ($A= 1, \ldots, 8$) are the generators in the fundamental 
representation of the SU(3) group.
In addition, there are also  chiral U(1) currents: 
\[
\mathcal{J} = :R^\dagger_{j\s}R_{j\s}:, \bar{\mathcal{J}} = :L^\dagger_{j\s}L_{j\s}:. 
\]
The U(1) symmetry does not correspond to electric charge which we do not treat here.
Instead the quarks carry baryonic charge. The interaction of 
the U(1) currents is an extra  feature absent in (3+1)-dimensions.
To analyze (\ref{model}) using non-abelian 
bosonization, we use the fact
that the Hamiltonian density of free Dirac fermions with symmetry U(1)$\times$ 
SU($N$)$\times$SU($M$) can be represented as a sum of a Gaussian U(1) theory and 
two Wess-Zumino-Novikov-Witten (WZNW) CFT models of levels $k=M$ and $k=N$ 
respectively \cite{knizhnik,book,affleckspinchain}. As a consequence, for each 
integer $n$ one can write down  $n$-point correlation functions of fermions in 
terms of products of $n$-point right and left conformal blocks of the U(1) 
Gaussian theory and the WZNW models. 

At $g >0$ the model (\ref{model}) is asymptotically free and acquires a mass gap 
$M_q = \Lambda g^{2/3}\exp(- 2\pi/3g)$ with $\Lambda$ being the ultraviolet 
cut-off, in the color sector even if the bare mass $m$ is zero. In the latter 
case the model  is integrable \cite{Ts86}. We consider this dynamically 
generated quark mass, $M_q$, as the largest energy scale in the problem.
The corresponding effective Lagrangian density for energies, $E \ll M_q$,  
is written in terms of the abelian and non-abelian Goldstone 
modes. It has a sigma model form \cite{gepner,bookqcd,book}: 
\begin{equation}
 {\cal L} = \frac{K}{2}(\p_{\mu}\theta)^2 + {\rm WZNW}[SU(2)_3;G] 
+m^*\mbox{Tr}(\re^{\ri\sqrt{2\pi/3}\theta}G + H.c.)  + 
\lambda\cos(\sqrt{8\pi/3}\theta),
\label{sigma}
\end{equation}
where $m^* \sim m$, $G$ is an SU(2) matrix corresponding to the 
WZNW field (see Appendix A), and the term labeled WZNW stands for  the WZNW Lagrangian on 
group SU(2) of level $k=3$.  In (\ref{sigma}), $\theta$ is a free massless 
bosonic field governing the U(1) density fluctuations and the Luttinger 
parameter $K$ is related to the abelian coupling $g_f$, so that at $|g_f| \ll 1$, 
$K -1 = O(g_f)$. For attractive interactions, $K <1$. 

There is a direct analogy with (3+1)-D case as a similar sigma model \cite{t'hooft} 
appears
there as an effective low energy action for QCD and is used to study mesons and baryons with the latter appearing as 
solitons.  The last term in Eqn. (\ref{sigma}), absent in the 
original formulation (\ref{model}), was introduced by t'Hooft \cite{t'hooft} who 
argued that instantons generate the term proportional to  the real part of 
the determinant of the U(2) matrix, $\exp(\ri\sqrt{2\pi/3}\theta)G$ 
(in condensed matter physics such a term is generated by Umklapp processes).

In Eqn. (\ref{sigma}), the SU(2) matrix WZNW field $G$ corresponds to the 
SU(2)$_3$ primary field 
$\Phi^{(j)}$ with $j=1/2$ and has scaling dimension 3/10 \cite{dms}. Therefore 
the scaling dimension of the interaction term in Eqn. (\ref{sigma}) with coupling 
constant $m^*$ is 
\[
d_{m^*} = \frac{1}{6K} + \frac{3}{10} ,
\]
so that it becomes irrelevant at $K < 5/51$.  We note however that this interaction
will generate at second order a relevant perturbation in the spin sector for $K<5/33$ so
that in fact the theory is always gapped in the isospin sector -- see section V.  The instanton term, $\cos 
(\sqrt{8\pi/3}\theta)$, in Eqn. (\ref{sigma}) has  scaling dimension $d_{\lambda} 
= 
2/(3K)$ and is relevant when $K > 1/3$. It becomes more relevant than the 
$m^*$-perturbation in Eqn. (\ref{sigma})  when $K > 5/3$.

 Below we list the most relevant  operators local in terms of quarks which 
survive after the projection onto the SU(3) singlet sector given by the ground 
state of the SU(3)$_2$ WZNW model perturbed by its current-current interaction 
(see 
 Appendix A). The operators with smallest scaling dimensions should correspond 
to   mesons (U(1) neutral particles with Lorentz spin zero which are two-body bound 
states of quarks) and baryons (particles with Lorentz spin 1/2 and 3/2, formed as
three-body bound states of quarks). 
 After projection  we obtain the following expressions for the meson operators 
at energy $E \ll M_q$: 
\be\label{mesonop}
\begin{split}
& {\vec M} = R^\dagger_{j\alpha} {\vec \s}_{\alpha\beta} L_{j\beta} \sim 
\re^{-\ri\sqrt{2\pi/3}\theta}\mbox{Tr}[\vec\s(G - G^\dagger)], 
\\
& M^0 = i ( R^\dagger_{j\alpha} L_{j \alpha}
- H.c.) \sim \ri\re^{-\ri\sqrt{2\pi/3}\theta}\mbox{Tr} \; G  + {\rm H.c.},
\end{split}
\ee
with $ {\vec \s}_{\alpha\beta}$ being the Pauli matrices.  The right-moving baryon 
operators given by the three-quark SU(3) singlet bound states with respective 
Lorentz spins 3/2 and 1/2 are: 
\bea\label{hadron}
&& \Delta_{3/2}^{\alpha\beta\gamma} = 
\epsilon^{abc}R_{a\alpha}R_{b\beta}R_{c\gamma} \sim 
\exp(3\ri\sqrt{2\pi/3}\varphi){\cal F}^{(3/2)}_{3/4},\label{H32}\\
&& n_{1/2}^{\alpha\beta\gamma} = \epsilon^{abc}R_{a\alpha}R_{b\beta}L_{c\gamma} 
\nonumber\\
&& \sim \exp[\ri\sqrt{2\pi/3}(2\varphi - \bar\varphi)]\left[{\cal 
F}^{(1)}_{2/5}\bar{\cal F}^{(1/2)}_{3/20}\right], \label{H12}
\eea
where $\varphi$ and $ \bar\varphi$ are the chiral components of the bosonic field, 
$\theta = \varphi +\bar\varphi$, and 
${\cal F}^{(j)}_{h_j}, \bar {\cal F}^{({\bar j} )}_{{\bar 
h}_{\bar j}}$) denote the SU(2)$_3$ 
holomorphic and anti-holomorphic conformal blocks with isospin $j, {\bar j}=0,1/2,1,3/2$ and 
weights $h_j = \frac{j(j+1)}{5}$.  Their counterparts with opposite 
chirality are given by similar expressions with $R$ replaced by $L$. 
As shown in Appendix A, the $\Delta_{3/2}$ operator 
has isospin $I=3/2$ (respectively $I=1/2$) and is called a $\Delta$-baryon 
in the following, while the $n_{1/2}$
operator has $I=1/2$ and is termed a nucleon.

There is also  a bosonic (Lorentz spin 0) dibaryon operator with isospin $I=0$ 
and U(1) charge $2$ (see Appendix A)  
which is the analogue of the deuteron. This operator is identified as follows in 
the low-energy $E \ll M_q$ limit:
\begin{equation}
d_0 = 
(R_{1\alpha}\epsilon_{\alpha\beta}L_{1\beta})(R_{2\gamma}\epsilon_{\gamma\delta}
L_{2\delta})(R_{3\eta}\epsilon_{\eta\rho}L_{3\rho}) \sim 
\exp(\ri\sqrt{6\pi}\phi)\mbox{Tr}(G + G^\dagger) ,
 \label{delta0}
\end{equation}
where $\phi = \varphi -\bar\varphi$ is the dual field to $\theta$. There is also 
a similar six-quark boson $\vec d$ with isospin $I=1$ described by Eqn. 
(\ref{delta0}) with $(G+G^\dagger)$ replaced by i$\s^a(G-G^\dagger)$.

\section{Semiclassical analysis of the  low-energy spectrum}

To get a qualitative understanding of model (\ref{sigma}) we can employ a 
semiclassical approximation  which will be later augmented by the numerical 
analysis based on the TCSA. Despite its {\it a priori} restricted validity, our numerical 
results presented in the next section confirm that this analysis presents a 
qualitatively correct picture of the excitation spectrum. 
 
To permit the semiclassical approximation, we represent 
the SU(2) matrix $G$ as $\hat G = \s\hat I + \ri \hat\s^a \pi_a, ~~ 
\s^2 + {\vec\pi}^2 = 1$.  Using this, we can write the interaction term in 
in (\ref{sigma}) as
\be
V = m^*\cos(\sqrt{2\pi/3}\theta)\sigma+
\lambda\cos(2\sqrt{2\pi/3}\theta).
\label{V}
\ee
The ground state is determined by minima of $V$ as the rest of 
the action (\ref{sigma})  contains derivatives of the fields.   The potential 
for $\lambda=0$ has degenerate minima at $\sqrt{2\pi/3}\theta=0, 2\pi
n,~ \sigma = -1$ 
and $\sqrt{2\pi/3}\theta = \pi(1+2n), ~~ \sigma=1$.  This fact suggests that there 
are two kinds of excitations: (i) fluctuations around the degenerate potential 
minima of  (\ref{V}) ($\eta$ and $\pi$ mesons); and (ii) kinks interpolating 
between these minima. The kinks interpolating between minima with 
the opposite sign of $\sigma$ correspond to baryons, while the ones interpolating between the 
vacua with the same sign of $\sigma$ are deuterons and isotriplet dibaryons. The 
small fluctuations constitute neutral isoscalar  ($\eta$) 
and isovector ($\pi$) mesons.  In Ref. \cite{t'hooft}
for  QCD3+1 it was predicted that there are eight mesons 
in total (one scalar and three vector mesons for each of the two vacua). In our numerics, we 
find these particles, but we also conclude that there are likely to be meson-meson bound states. 

We have discovered that our numerical calculations
indicate that we can make our semiclassical analysis quantitative.
More precisely, the numerical calculations demonstrate  that for $0.3 < K < 1.5$ and $\Lambda =0$ the 
charge sector is well described by the sine-Gordon model. 
Hence for analytical calculations one can use  the decoupling procedure where the bosonic exponent in Eq. 
(\ref{sigma})  by its average and that similarly the U(1) charge sector 
can be well described by replacing Tr$G$ by its vacuum average
\bea \label{appr}
m^*\mbox{Tr}(\re^{\ri\sqrt{2\pi/3}\theta}G + {\rm H.c.}) &\rightarrow &
m^*(\langle\re^{\ri\sqrt{2\pi/3}\theta}\rangle \mbox{Tr}G + 
\re^{\ri\sqrt{2\pi/3}\theta}\langle\mbox{Tr}G\rangle + {\rm H.c.}),
\eea
where the theory that then describes the $U(1)$ boson is sine-Gordon
with coupling proportional to $\langle \sigma \rangle$.   In this
decoupling
procedure, the scalar mesons correspond to the sine-Gordon breathers and
the deuterons to the sine-Gordon solitons.
As is known, the breathers disappear from the 
spectrum when the scaling dimension of $\cos(\sqrt{2\pi/3}\theta)$ becomes  
larger than $1$ corresponding to $K <1/6$. For $K >1/6$ there are bound 
states of the first breather corresponding to bound states of the scalar 
mesons. 
If we assume that this procedure continues to work for $\lambda \neq 0$, 
it will yield the double sine-Gordon model for the 
charge sector.

We can understand the effects of a nonzero $\lambda >0$ if we expand (\ref{V}) 
around a particular minimum.  For the sake of argument, we suppose $m^*\sigma>0$:
\be
V \sim  A(\theta - \sqrt{3\pi/2})^2 + B(\theta - \sqrt{3\pi/2})^4,
\label{eff}
\ee
where $B>0$ and $A$ may change sign depending on the mutual strength of 
$m^*$ and $\lambda$. For $A>0$ the minimum is located at $\theta = \sqrt{3\pi/2}$. 
At $A<0$ the minima shift away from these points and $A=0$ corresponds to the 
Ising phase transition as in the double  sine-Gordon model 
\cite{mussardo,nersesyan,dsgtcsa}.  The transition occurs 
when $\lambda^{1/(2-d_{\lambda})} \sim {m^*}^{1/(2-d_{m^*})}$, that is at
 $\lambda \sim  (m^*)^{4(3K-1)/(51K/5 -1)}$ 
and $K >1/3$, i.e. when both operators in Eqn. (\ref{V}) are 
relevant. When $\lambda$ exceeds the critical value  the minima of the potential 
(\ref{V}) split so that the effective potential (Eqn. \ref{V}) has two minima in 
the unit cell $0< \sqrt{2\pi/3}\theta < 2\pi $. Then the bosonic vertex operator $\sin(\sqrt{2\pi/3}\theta)$ 
acquires a nontrivial vacuum expectation value. 
This implies that $M_0$ (\ref{mesonop}) also has a non-trivial vacuum expectation value.
This Ising order parameter having a finite expectation value might suggest that 
the condensate of mesons with zero SU(2) isospin is topologically nontrivial in the sense
of Kitaev \cite{kitaev}.  However because this Ising order is occurring in an interacting
system, we cannot conclude definitively that this phase is topological.
In the condensed matter context a similar 
spontaneously dimerized phase emerges when one increases the electron-electron 
repulsive interaction in an insulator \cite{nersesyan}.  The splitting of the 
minimum of the effective potential creates a possibility that U(1) charged 
particles with isospin zero (deuterons) will have two different U(1) (baryon) 
charges corresponding to the short and long kinks of the field $\theta$. 

When $\lambda <0$ we point to the possibility that two kinks (not kink-antikink)
of the double sine-Gordon model may have bound states because an attractive potential in 1D
generically leads to bound states.  Such states would correspond to nuclei with baryonic charges higher than that of 
deuterium.   However we will leave the resolution of this question to future work.

\section{Non-perturbative numerics}  

To go beyond the semiclassical approximation and to determine the coherence and 
stability of the low-energy excitations we have investigated the 
non-perturbative energy spectrum of model (Eqn. \ref{sigma}) by TCSA. The 
bosonized form (\ref{sigma}) permits a straightforward application of TCSA using 
its recent extension to treat deformations of conformal field theories of the 
WZNW-type \cite{konik}.

In the case $\lambda=0$, $m \neq 0$ and zero particle density we have found that 
the operators (\ref{mesonop}) and (\ref{delta0}) indeed annihilate (create) 
coherent single-particle excitations. In particular $d_0$ and $\vec d$ destroy 
coherent excitons with isospin $I=0$ and $I=1$ respectively, with masses below 
the baryon continuum. Masses of the nucleons, mesons, and deuterons determined 
from the TCSA for different values of $K$ are listed in Table \ref{tabmasses}. The 
mass ratios of the isoscalar particles turn out to be well described by  the 
known exact result for the sine-Gordon model (see Eqn. \ref{SGmass}), which is an 
important evidence for the validity of the approximation in Eqn. \ref{appr}).
For the case where $\lambda$ is nonzero, our numerics clearly indicate a phase transition at 
finite positive $\lambda$.

Below we discuss the application of TCSA to the model (\ref{sigma}) and present 
a detailed analysis of the numerical results.

\renewcommand{\arraystretch}{1.2}

\begin{table}[!b]
\begin{center}
\begin{ruledtabular}
    \begin{tabular}{ l  r  l  l  l  l  r}
    particle species&$K=0.4$&$0.6$&$0.8$&$1.0$&$1.2$&$1.4$\\
    \hline
    nucleon\footnote{The mass of the lightest baryon is estimated from the 
two-particle continuum in the baryonic charge two 
sector.}&$4.5$&$4.3$&$4.5$&$4.8$&$5.1$&$5.4$\\
    isoscalar meson&$5.5$&$3.9$&$3.2$&$2.8$&$2.5$&$2.3$\\
    isovector meson&$3.6$&$3.1$&$2.9$&$2.8$&$2.7$&$2.7$\\
    isoscalar deuteron&$6.7$&$7.4$&$8.2$&$8.9$&$9.7$&$10.2$\\
    isovector deuteron&$8.2$&$8.3$&$8.7$&$9.2$&$9.7$&$10.2$\\
    \end{tabular}
\end{ruledtabular}
    \end{center}
\caption{Masses of the low-energy particles at $\lambda=0$ and zero particle 
density determined from TCSA in units $M=(m^*)^{1/(2-d_{m^*})}$. We estimate the 
error to be $0.5\,M$ and $1\,M$ in the meson and deuteron sectors, respectively, 
independent of $K$, and a relative accuracy to be one order of magnitude 
smaller.}\label{tabmasses}
\end{table}

\subsection{The settings of the TCSA}
Following standard procedures of TCSA 
(see e.g. \cite{truncated,dsgtcsa,konikadamov,konik}), the quantum field theory 
is 
considered in a periodic spatial box of volume $L$. Introducing a unit of 
mass $M$ as 
\begin{equation}
m^*=M^{2-d_{m^*}}\qquad d_{m^*}=\frac{1}{6K}+\frac{3}{10} 
\end{equation}
we can define a dimensionless volume parameter $l=ML$ and measure energies in 
units of 
$M$. Mapping the coordinates $(\tau,x)$ (where $\tau$ is Euclidean time) to the 
complex 
plane 
\begin{equation}
z=\re^{\tau-\ri x}\; , 
\end{equation}
 the dimensionless Hamiltonian can then be written as
\begin{equation}
h(l)=\frac{1}{M}H(L)=\frac{2\pi}{l}\left(L_{0}+\bar{L}_{0}-\frac{c}{12}
\right)+\frac{l^{1-d_{m^*}}}
{(2\pi)^{-d_{m^*}}}B_{m^*} 
+M^{d_\lambda-2}\lambda\frac{l^{1-d_{\lambda}}}{(2\pi)^{-d_{\lambda}}}B_{\lambda
},
\end{equation}
where $L_{0}+\bar{L}_{0}-c/12$ is the conformal term and $B_{m^*}$, $B_\lambda$ 
are 
the matrices of the operators
\begin{equation}
\cos\left(\sqrt{2\pi/3}\theta(z,{\bar 
z})\right)\left(\Phi_{-1/2,1/2}^{(1/2)}(z,{\bar z})
-\Phi_{1/2,-1/2}^{(1/2)}(z,{\bar z})\right)
\end{equation}
and
\begin{equation}
\cos\left(2\sqrt{2\pi/3}\theta(z,{\bar z})\right),
\end{equation}
respectively, at the point $z=1$ in the basis spanned by the conformal states. 
This space is truncated 
by keeping states for which 
\begin{equation}
L_{0}+\bar{L}_{0}<N_{cut},
\end{equation}
for some cutoff value $N_{cut}$.  With the cutoff in place, the resulting Hamiltonian can be 
numerically 
diagonalized. Due to translational invariance, the space can be split into
sectors corresponding to the value of momentum; we restrict ourselves to 
discussing the zero-momentum sector, sufficient to obtain the information 
we need. In addition, it is possible to split the Hilbert space according to
the value of baryonic charge and the third component of isospin, which makes 
the computation more efficient. 

The matrix elements of the perturbing
operators can be evaluated by reducing them
to conformal field theory structure constants
using conformal Ward identities (for example \cite{kormos}) for an explicit description of the
procedure which can be easily adapted
to the Kac-Moody algebra considered here).
Structure constants for diagonal
primary fields in the SU(2) WZNW models
have been obtained in \cite{FZ}, while those of
the bosonic part are trivially obtained from
free field theory.
The inherent cut-off dependence is reduced 
using 
NRG techniques \cite{konikadamov} to push $N_{cut}$ to higher values, and 
analytic RG 
techniques to eliminate the cut-off dependence by introducing 
a running coupling following the procedure in 
\cite{lencses,rychkov1,rychkov2,watts}.
We performed calculations in the parameter regime $0.3\leq K\leq1.5$ with a 
step size of 0.1. Representative spectra are shown and analyzed below.

\subsection{Mesons}

\begin{figure}
\begin{centering}
\includegraphics[scale=0.4]{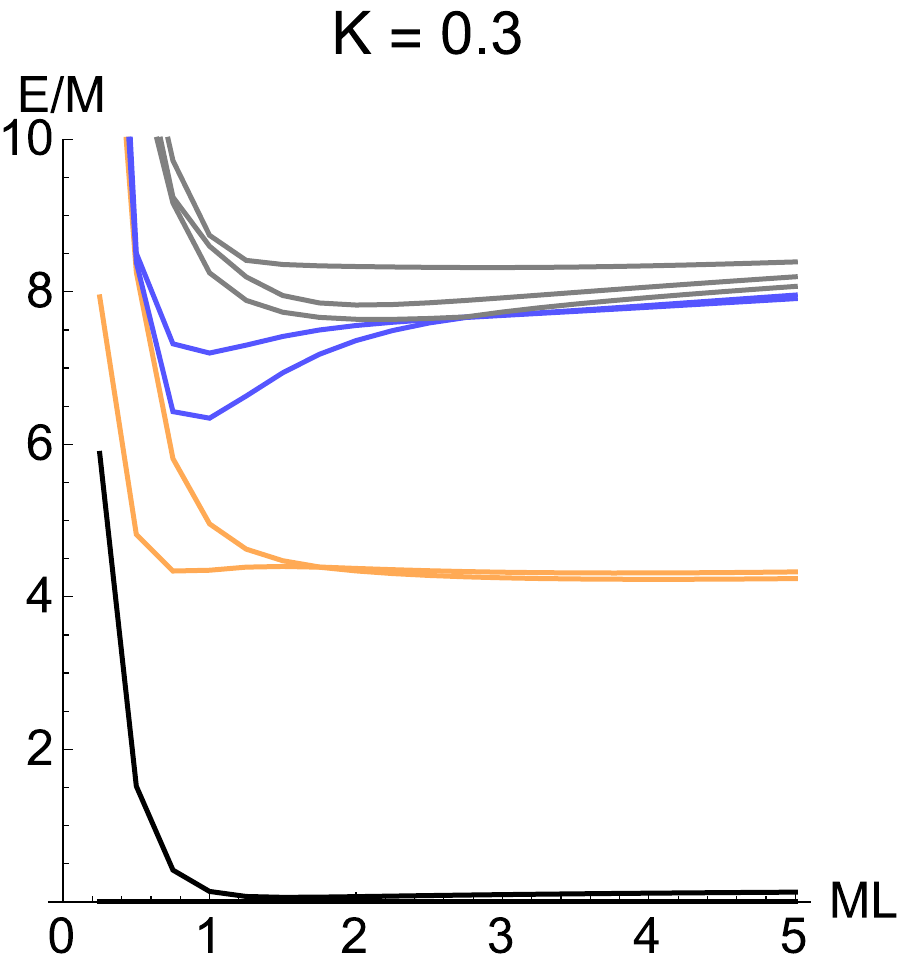} \includegraphics[scale=0.4]{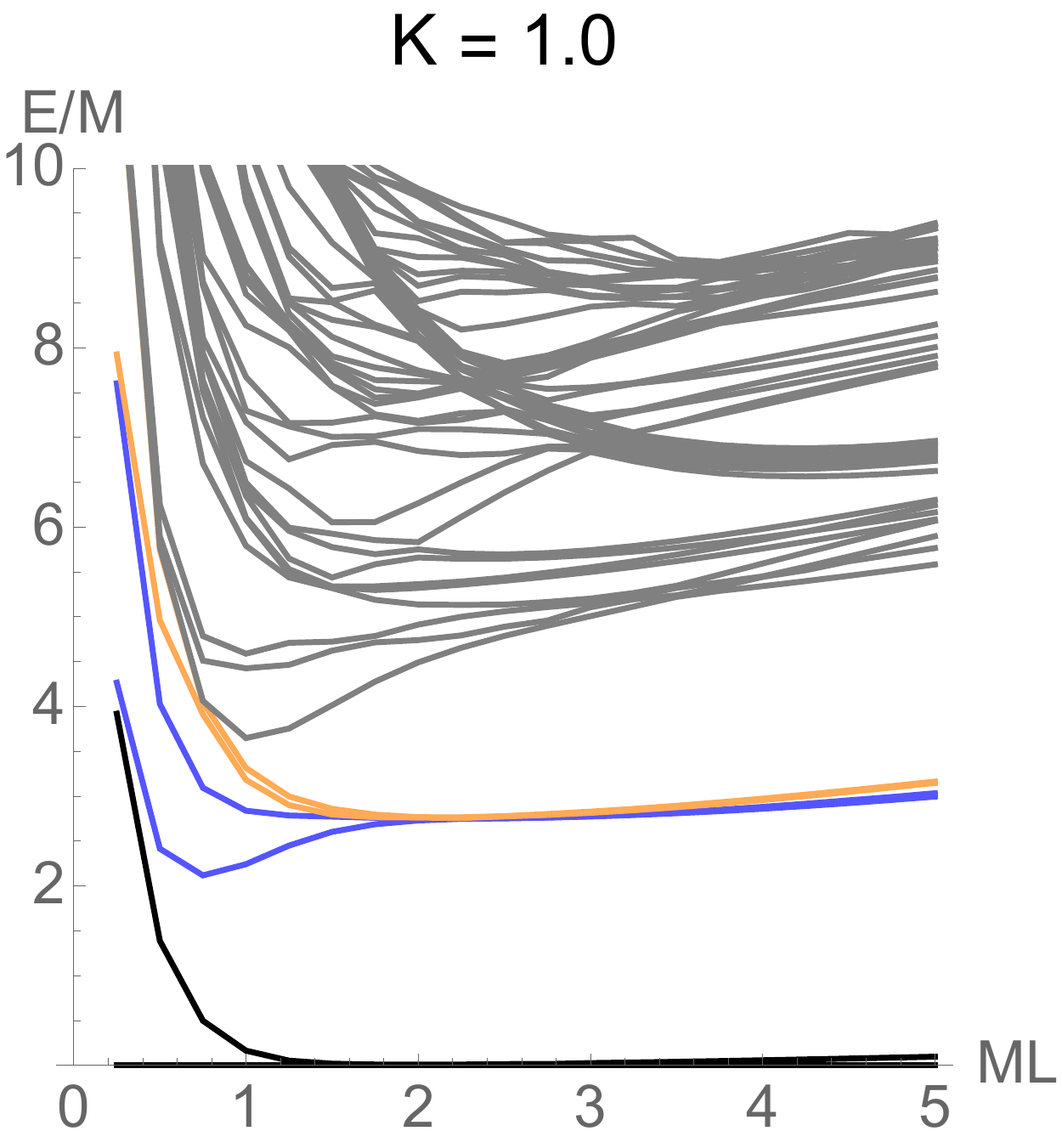}
\includegraphics[scale=0.4]{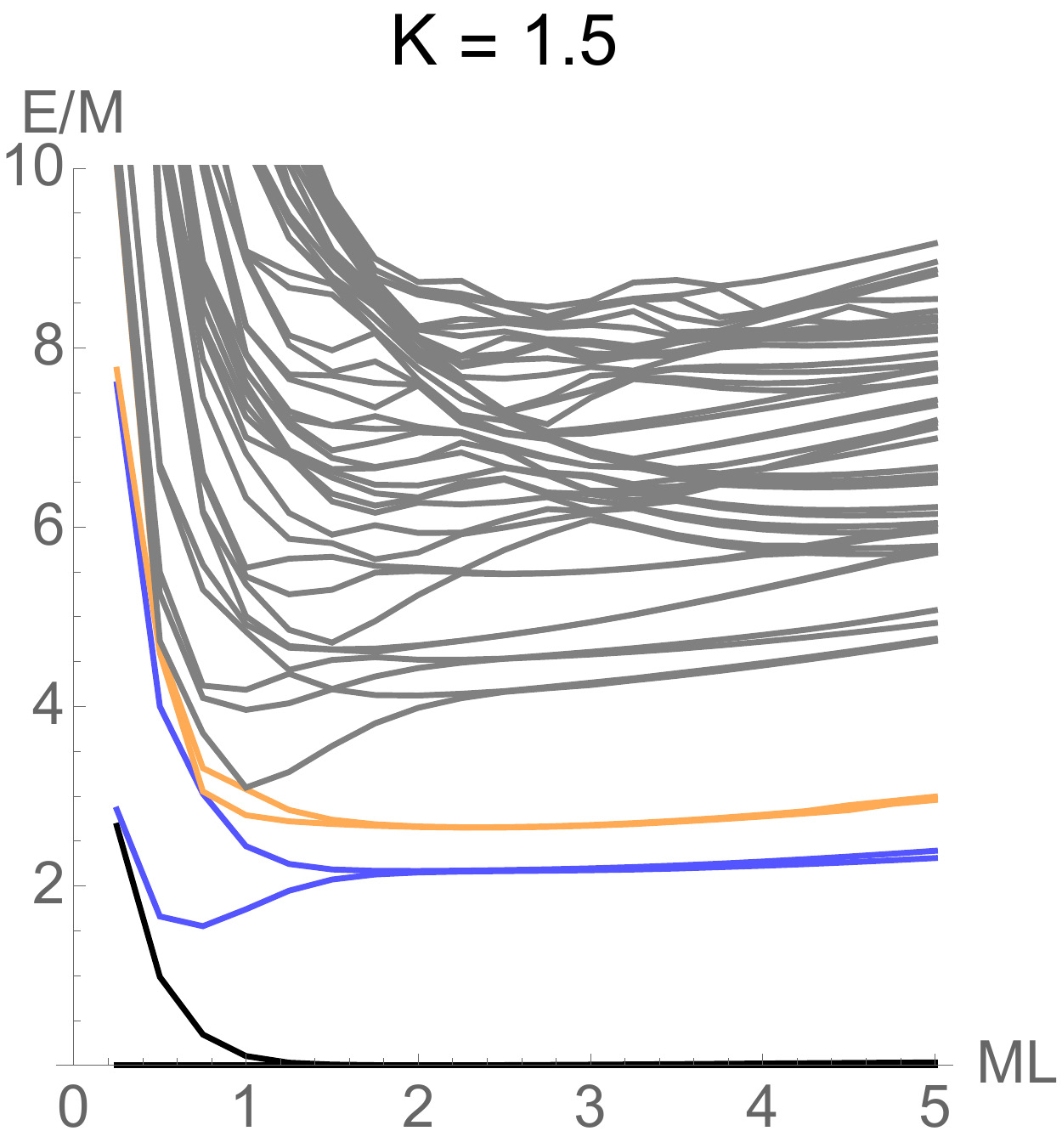}
\par\end{centering}
\caption{Finite volume spectra in baryon charge zero sector at $K=0.3$, 1.0, 
1.5. The vacua (black), and the mesons (isosinglets -- blue, isovectors -- 
orange) are highlighted.}
\label{fig:mesonspectra}
\end{figure}

Initially we study the $\lambda=0$ case. We plot the zero momentum levels in the 
zero baryonic charge sector in 
Fig. \ref{fig:mesonspectra} for the values $K=0.3$, $1$ and $1.5$, 
with levels colored according to the isospin multiplet they belong to.
The spectra are normalized by subtracting the ground state. There is another 
vacuum state which becomes exponentially degenerate with the lower 
ones for large volume, and accordingly all neutral particles come in 
two copies. 

One-particle states can be identified by being below and separated 
from the dense continuum, and tending exponentially to a flat behavior. 
Some of these levels  also show a dip before becoming flat for larger 
volumes, which is a behavior that is typical for one-particle states 
due to so-called $\mu$-terms dominating their finite size corrections
\cite{luscher}.

We find a meson isotriplet ($\rho$) and an isosinglet 
($\omega$) meson, two copies for each. For $K=1$ the numerics shows 
them to be degenerate.  When $K>1$ the triplet is heavier than the singlet, 
while 
for $K<1$ it is the mass of the singlet that is heavier than that of triplet.

\begin{figure}
\begin{centering}
\includegraphics[scale=0.6]{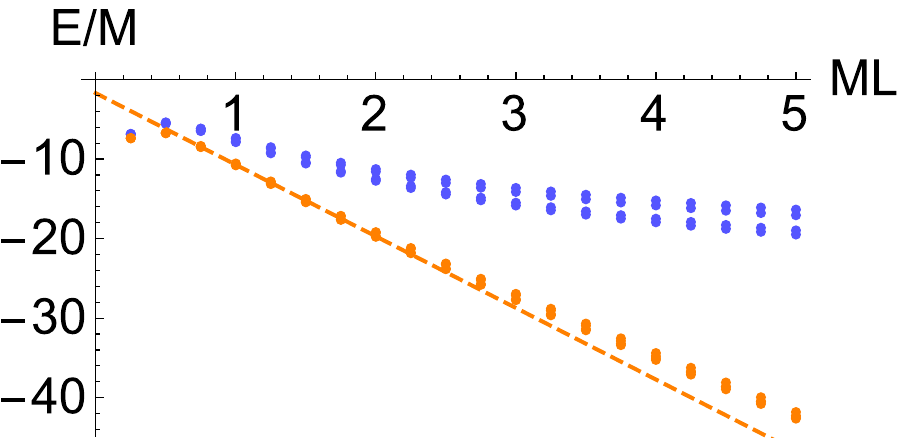} \includegraphics[scale=0.6]{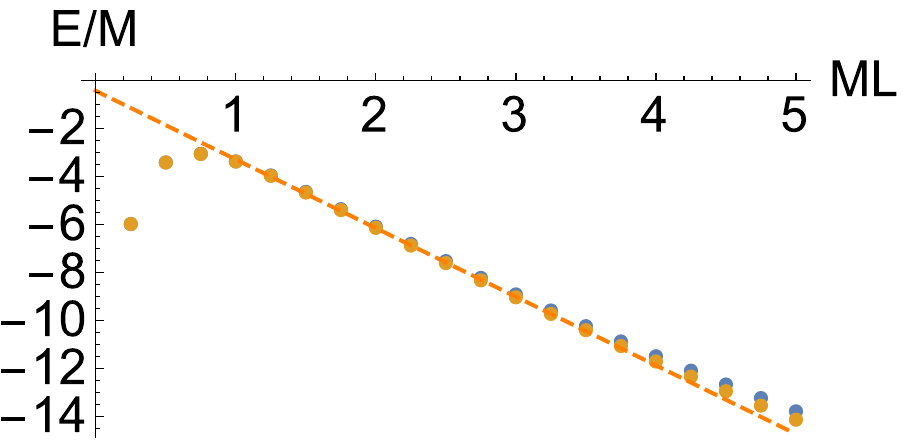} 
\par\end{centering}

\begin{centering}
\includegraphics[scale=0.6]{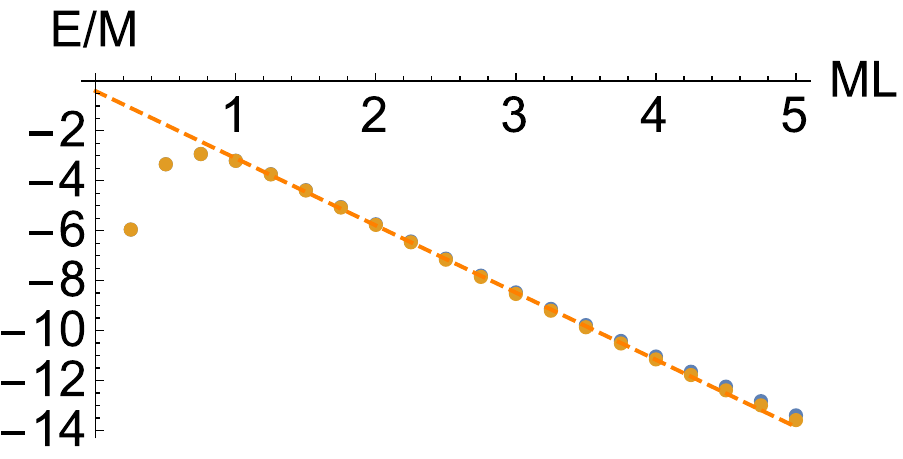} \includegraphics[scale=0.6]{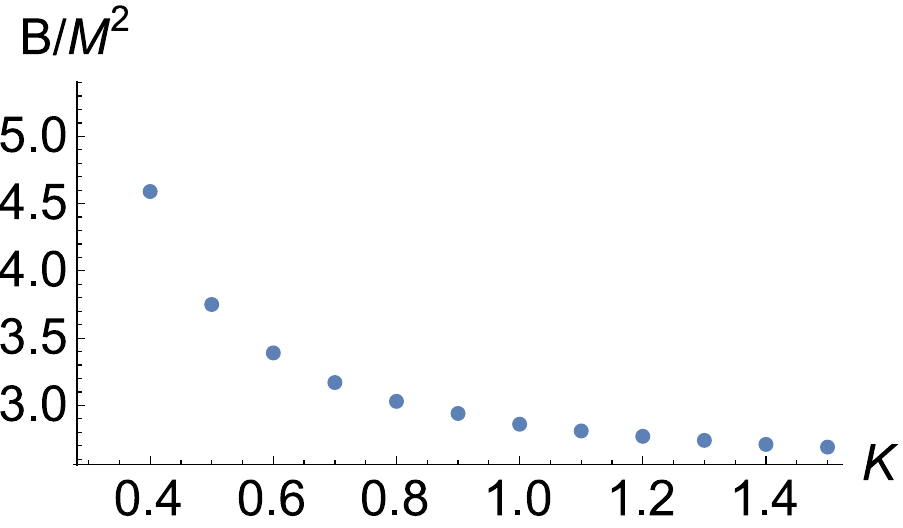} 
\par\end{centering}

\caption{Raw (blue) and renormalized (orange dots) vacuum energies for $K=0.3$,
1.0, 1.5 and at energy truncation $E_{cut}/M=8$. The lines are linear
fits yielding the energy densities in the bottom right panel.}
\label{fig:bulk}
\end{figure}

From the ground state level we can also measure the bulk energy density, as 
shown in Fig. \ref{fig:bulk}. For this we must perform a perturbative 
renormalization improvement to eliminate the leading dependence 
on the cutoff $N_{cut}$; we follow the procedure used in \cite{konik}. 
This improvement leads to scaling the levels corresponding to different 
cut-offs on top of each other, and is more relevant for smaller $K$. 
The resulting vacuum energy densities $B$ defined by the asymptotic 
behaviour
\[
E_{0}(L)=-BL+\ldots
\]
where the omitted terms decay exponentially in the volume. The results of 
the linear fit are shown in the fourth subfigure of Fig. \ref{fig:bulk}.

For excited states, we implemented further RG corrections corresponding 
to introducing a running coupling as in \cite{lencses}, however for this model 
the corrections induced by this proved to be negligible.  We 
also added cut-off extrapolation improvements following \cite{lencses}. 
Results from even and odd cut-offs can be extrapolated separately, and
the extrapolated results are consistent, but they do not lead to any 
appreciable improvement of the results either.

\begin{figure}
\begin{centering}
\includegraphics[scale=0.8]{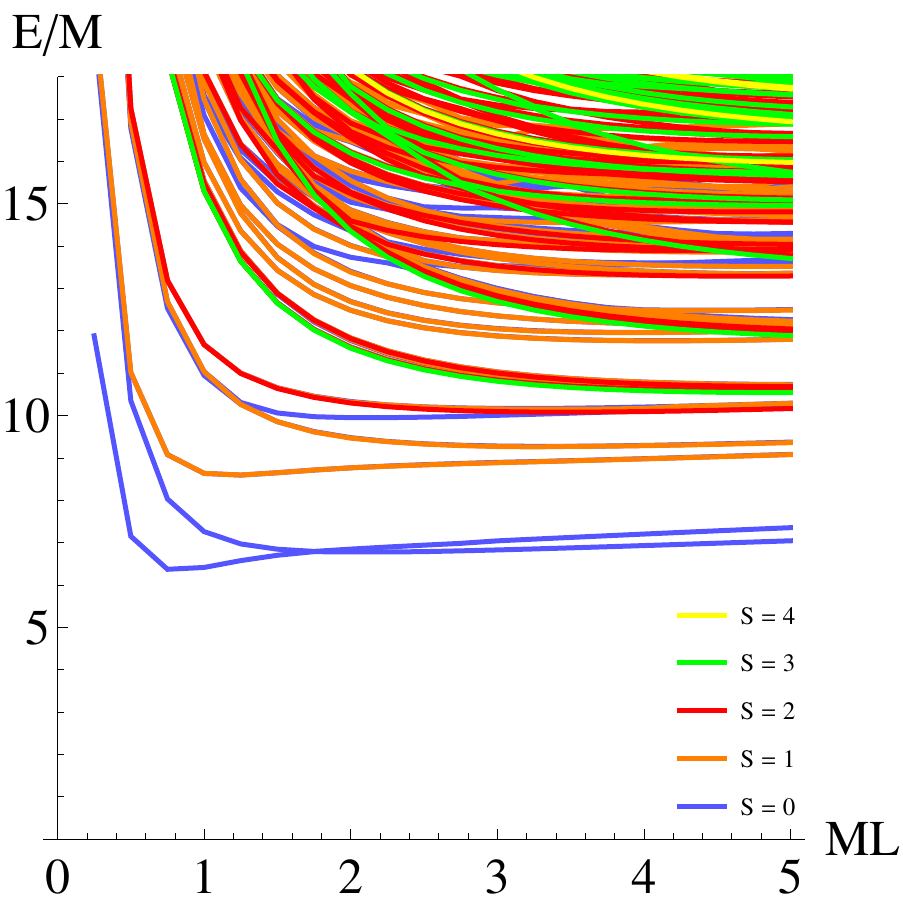} 
\par\end{centering}

\caption{Finite volume spectra for $K=0.3$ in the baryonic charge 2 sector, also 
showing the different isomultiplets formed by the various lines.}
\label{fig:dibaryons}
\end{figure}

\subsection{Baryon bound states}

Presently, the truncated conformal space approach
can only access sectors of even baryonic charge. Sectors
of odd baryonic charge correspond to half-integer
isospin, i.e. transform under spinorial representations
of the diagonal subgroup of the SU(2)$_L$ x SU(2)$_R$
symmetry. Such states can only arise from fields that
transform differently under the left and right SU(2)
groups. However, the structure constants of such
fields are presently unknown; only the ones
corresponding to diagonal fields are available \cite{FZ}.
In the sector with baryon charge $2$ we find both isospin singlet 
and triplet bound states, as shown by the spectrum in Fig. \ref{fig:dibaryons}. 
The isospin singlet state (coming in two copies due to the doubly degenerate 
vacuum is the analogue of the deuteron. 
However, the isotriplet has 
no analogue in 3+1 dimensional strong interactions; the
reason is that in the isospin triplet channel the nucleon-nucleon potential 
is not attractive enough to form a bound state. However in one-dimensional
space any attractive potential -- no matter how shallow -- leads to a bound
state. From these considerations the appearance of the isotriplet bound state 
(two members of this multiplet would correspond to proton-proton and neutron-neutron bound states
in the usual language of the strong interaction) is not so surprising. For higher values of 
$K$ one of the triplets joins the continuum, 
i.e. the corresponding bound state disappears.

\begin{figure}
\begin{centering}
\includegraphics[scale=0.8]{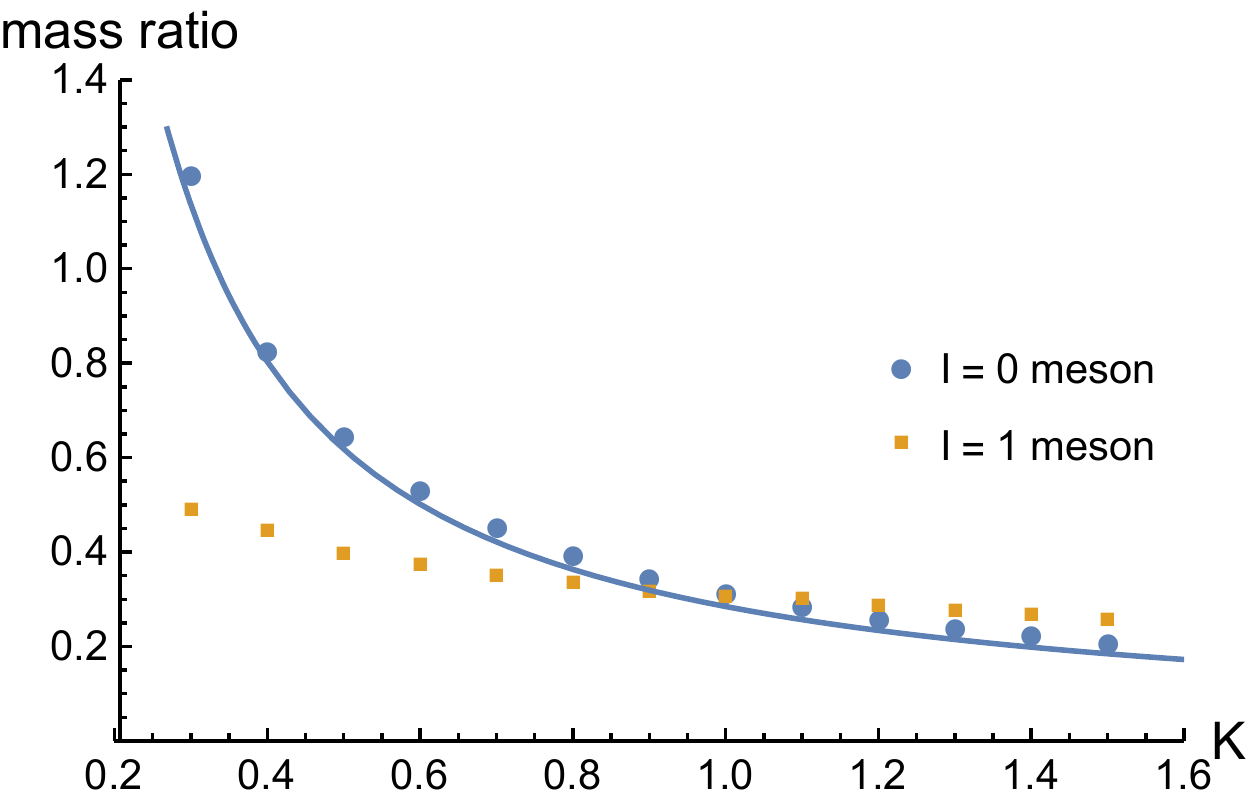}
\par\end{centering}
\caption{Isoscalar and isovector meson/dibaryon mass ratios ($m_{\rm meson}/m_{dibaryon}$) and the approximate estimate
for this in Eqn. \ref{SGmass}.}
\label{fig:mdbratio}
\end{figure}

Assuming the validity of the semiclassical decoupling in Eqn. \ref{appr}
one expects the dynamics of isosinglet states will be dominated by the sine-Gordon like 
dynamics of the field $\theta$ as the isospin degrees of freedom corresponding 
to the WZNW field $G$ are frozen. The deuteron excitations 
correspond to sine-Gordon solitons, while the isosinglet mesons to the 
breathers. 
Their masses are known to be related by the exact relation
\be
\frac{m_{\text{meson}}}{m_{\text{dibaryon}}}=2\sin\left(\frac{\pi/2}{12K-1}
\right), \label{SGmass}
\ee
which is comparable to the numerically observed mass ratio in Fig. 
\ref{fig:mdbratio} so providing numerical support 
for the decoupling approximation in Eqn. (\ref{appr}).
Our plot also shows that this ratio does not hold in the isotriplet channel, 
which can be expected since these excitations have a more complicated 
structure due to the presence of the isospin degrees of freedom.

\subsection{The case $\lambda\neq 0$}

\begin{figure}
\begin{centering}
\includegraphics[scale=0.55]{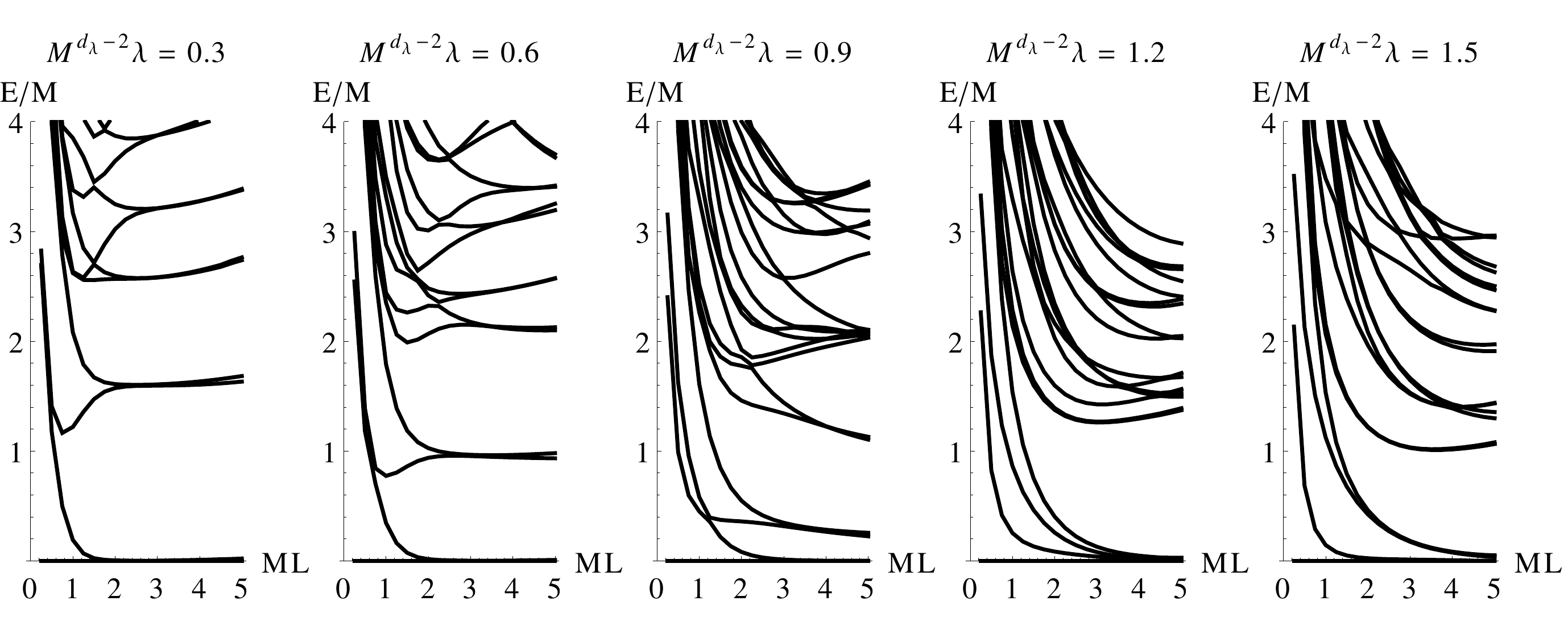}
\par\end{centering}
\caption{Finite volume spectra at $K=1.5$ and different settings of the 
dimensionless coupling $M^{d_\lambda-2}\lambda$ in baryonic charge zero sector. 
$E_{cut}/M=8$.}
\label{fig:lambda}
\end{figure}

In Fig. \ref{fig:lambda} we show spectra at fixed $K=1.5$ and with increasing 
$\lambda$. The most prominent feature is a change in the vacuum structure: two 
additional vacua develop from the lines corresponding to the $\lambda=0$ singlet 
mesons. This supports the existence of the critical point established by our 
semiclassical analysis. For $K=1.5$ we estimate the critical value of the 
coupling to be $M^{d_\lambda-2}\lambda=1.2$. Above this value the vacua come in 
two pairs approaching each other algebraically.
With increasing $\lambda$ the triplet mesons join the continuum.  We conclude therefore in 
the large $\lambda$ phase that all eight mesons are unstable.

\section{ Finite baryon density} 

New phases emerge when considering the model at finite baryon   
density by introducing into the Hamiltonian (\ref{model}) a chemical potential, 
$\mu$, coupled to the baryonic charge (recall that the baryonic charge of quarks is 1/3)
\bea
V_{\mu} = \mu(2/3\pi)^{1/2}\p_x\theta.
\eea
This explicit chemical potential term can be removed using the position 
dependent field redefinition $\sqrt{2\pi/3}\theta \rightarrow 
\sqrt{2\pi/3}\theta + 2k_Fx$, where the Fermi vector is $k_F \sim (\mu^2- 
M_n)^{1/2}$, with $M_n$ being the nucleon mass. Since  the chemical potential 
has no influence on the color sector, the ground state remains a color singlet 
and the quarks remain massive. 

At finite baryon density the terms containing exponents of $\theta$ field in the 
effective Lagrangian (\ref{sigma}) become oscillatory. Normally such
terms are dropped from the Lagrangian.  However at second order in perturbation theory,
the $m^*$-term coupling $SU(2)_3$ with the $U(1)$ boson may
give rise to the term in the action
\begin{equation}
- A(\eta) (m^*)^2  \mbox{Tr}\Phi^{(1)}(x,\tau), \label{Perturb}
\end{equation}
where $\Phi^{1}$ is the spin-1 SU(2)$_3$ primary field.  At second order in perturbation theory in $m^*$,
this field is generated by the operator product of the ${\rm Tr}G$ fields:
\bea
&& \re^{\ri(\sqrt{2\pi/3}\theta(x_1,\tau_1) + 2k_Fx_1)}\mbox{Tr}G(x_1,\tau_1)\re^{-\ri(\sqrt{2\pi/3}
\theta(x_2,\tau_2) + 2k_Fx_2)}\mbox{Tr}G(x_2,\tau_2)  = \nonumber\\
&& \re^{2\ri k_F(x_1-x_2)}{(x_{12}^2 + \tau_{12}^2)^{-d_{m^*}+d_{adj}/2}}\mbox{Tr}\Phi^{(1)}(x_2,\tau_2) +...
\eea
Here  $d_{m^*} = 1/6K +3/10$ and $d_{adj} = 4/5$.
The operator (\ref{Perturb}) is generated if $A(\eta)$, the coefficient of the operator in the effective action given by,
\begin{equation}
A(\eta) =
\int_{\Lambda_{eff}} \rd\tau\rd y \frac{\cos(2k_Fy)}{(y^2 + \tau^2)^{(\eta+1)/2} },
\end{equation}
converges at large distances where $\eta = 2d^*_m-d_{adj}-1$.   This integral comes equipped with an effective UV cutoff, $\Lambda_{eff}$, 
that reflects we have dropped all higher order terms in the OPE.  $\Lambda_{eff}$ is proportional
to $1/R$ where $R$ is the size of the region where the integrand
$$
\frac{\cos(2k_Fy)}{(y^2 + \tau^2)^{(\eta+1)/2}},
$$
in the above integral has significant support and encodes the fact that ${\rm Tr}\Phi^{(1)}(x)$ is
really smeared over a region $R$.  
As $K \rightarrow K_c$, $R$ goes to $\infty$ and $A(\eta)$
vanishes.  For $K$ away from $K_c$, we can take $\Lambda_{eff}$ to $\infty$ and explicitly
evaluate $A(\eta)$:
\begin{equation}
A(\eta) = \int_0^{\infty} \rd r J_0(2k_Fr)r^{-2d_{m^*}+d_{adj} +1} \sim k_F^{\eta-1}\frac{\Gamma(1/2-\eta/2)}{\Gamma(1/2+\eta/2)},
\end{equation}
This integral is IR convergent if $\eta= 1/3K - 6/5 > -1/2$
or $K < K_c = 10/21$.   Note the effect that finite baryon density, i.e. $k_F \neq 0$, has.
At zero density, the integral is instead IR convergent if $\eta > 1$ which corresponds to $K < 5/33$.

Returning to finite density case, we furthermore can estimate $R$ to be
\begin{equation}
R = \bigg(\sqrt{\frac{\pi}{2}}\frac{2^{-\eta}\Gamma(\frac{1-\eta}{2})}{\Gamma(\frac{1+\eta}{2})}\bigg)^{-\frac{1}{\eta+1/2}}.
\end{equation}

The effective action for $K < K_c = 10/21$ is then
\begin{equation}
 {\cal L} = \frac{\tilde K}{2}\Big[\frac{(\p_{\tau}\theta)^2}{v_c} + 
v_c(\p_x\theta)^2\Big]  
 +{\rm WZNW}[SU(2)_3;G]  - \gamma \mbox{Tr} \Phi^{(1)},\label{new}
\end{equation}
Here $\tilde K$ and the Fermi velocity, $v_c$,
depend on the bare $K$ and the chemical potential, $\mu$.  $\gamma$ is
defined as
$$
\gamma = A(\eta)(m^*)^2 .
$$
Because $A(\eta)$ vanishes as $K\rightarrow K_c$, we see that the mass generated by the
relavent perturbation, $\mbox{Tr} \Phi^{(1)}$, also goes to zero.

\subsection{ Baryon metal phase, $K > K_c$} 

Hence we have two different situations depending on the strength of the 
attractive forward scattering in the original model. For sufficiently weak 
attractive interactions, $K> K_c$, the adjoint operator, $\Phi^{(1)}$, is not generated
Then at distances $\gg k_F^{-1}$  the finite particle density phase is described by two 
critical models: 
\begin{equation}
 {\cal L} = \frac{\tilde K}{2}\Big[\frac{(\p_{\tau}\theta)^2}{v_c} + 
v_c(\p_x\theta)^2\Big]  
 +{\rm WZNW}[SU(2)_3;G]  ,\label{baryonmetal}
\end{equation}
i.e. the  U(1) Gaussian model and the SU(2)$_3$ WZNW model 
describing the $\pi$-meson field. Their spectra are linear with velocities 
$v_c$ and $v_s$ respectively. 
Due to the vanishing gaps, this critical phase is a conductor 
that we call a baryon metal. The asymptotics of the 
correlation functions of the baryon operators  can be extracted from the 
identification (\ref{H32},\ref{H12}) and we find:
\begin{eqnarray}
\la \Delta_{3/2}(\tau,x)\Delta_{3/2}^\dagger(0,0)\ra &=&  \frac{Z_{3/2}\Big(\frac{\tau_0^2}{\tau^2 
+(x/v_c)^2}\Big)^{\tilde\eta_{c,3/2}}  }{[(\tau - \ri x/v_c)(\tau - \ri 
x/v_s)]^{I}};\cr\cr
\la n_{1/2}(\tau,x)n_{1/2}^\dagger(0,0)\ra &=&  \frac{Z_{1/2}\Big(\frac{\tau_0^2}{\tau^2 
+(x/v_c)^2}\Big)^{\tilde\eta_{c,1/2}}\Big(\frac{\tau_0^2}{\tau^2 
+(x/v_s)^2}\Big)^{\tilde\eta_{s,1/2}}   }{[(\tau - \ri x/v_c)(\tau - \ri 
x/v_s)]^{I}},\label{HH}
\end{eqnarray}
where $\tau_0 \sim K_F^{-1}$ is an ultraviolet cutoff and  
${\tilde \eta}_{c,3/2} = \frac{3}{8}(\sqrt {\tilde K}  - 1/\sqrt {\tilde K})^2$, 
${\tilde \eta}_{c,1/2} = \frac{1}{24}(1/\sqrt {\tilde K} - 3 \sqrt  {\tilde K} )^2$, 
and $\tilde\eta_{s,1/2}=3/10$.
The correlator of the left-moving particles is obtained by $x \rightarrow -x$  
and the one of the right- and the left- moving ones is zero.

From Eqn. (\ref{HH}) one can see that the baryons are incoherent:
their single particle Green's functions have branch cuts which is the
hallmark of a Tomonaga-Luttinger liquid \cite{bookboso}. In such a liquid there 
are no quasiparticles possessing both isospin and baryonic charges.  Instead 
they decay into cascades of collective excitations ($\pi$-mesons) propagating 
with different velocities.

In this phase, there are instabilities to both deuteronic superconductivity and a meson
density wave.
The instability towards deuteronic superconductivity 
is revealed as a singularity in the response function of the deuteron operator 
(see Eqn. (\ref{delta0})).  The scaling dimension of this field is 
$d_{deut} = 3(\tilde K/2+1/10)$, yielding a singular susceptibility as $T \rightarrow 0$ for $\tilde K \leq 7/15$:
\bea
\chi_{deut} = \int_0^{1/T}\rd\tau\int \rd x \la\hat T d_0(\tau,x) 
d_0^\dagger(0,0)\ra \sim T^{-2 +2d_{deut}},
\eea
where $T$ is temperature.  We note that this instability can be in either the isospin
$I=0$ or $I=1$ channels.  

The meson density wave instability (with wavevector $2k_F$) is, on the other hand, 
related to the scalar meson operator  
$M_{0}= R^\dagger_{a\alpha}L_{a\alpha} \sim \re^{-\ri\sqrt{2\pi/3}\theta}\mbox{Tr}G$.
This scaling dimension of this order parameter is 
$
d_{DW} = \frac{1}{6\tilde K} + \frac{3}{10} .
$
This order becomes singular at low temperatures for $\tilde K \geq 5/21$.

\subsection{Phase with gapped isospin excitations, $K < K_c$} 

At $K <K_c$ the effective action (\ref{new}) contains the relevant perturbation, $\gamma {\rm Tr}\Phi^{(1)}$. 
Since the operator $\Phi^{(1)}_{m,\bar m}$ acts only in the 
isospin sector the field $\theta$ remains gapless with central charge $c=1$. The 
SU(2) sector of  (\ref{new}) has been studied in \cite{konik,affleckhaldane}.
At $\gamma >0$ (the present case) it becomes massive and the spectrum consists  of 
two degenerate massive triplets (pions).  These baryons 
are incoherent, i.e. their correlation functions do not have poles in frequency-momentum plane,  only branch cuts (see Appendix C).

Like for $K>K_c$, here there is an instability to deuteronic superconductivity.
Unlike in the baryon metal phase at $K < K_c$, the amplitude of the $SU(2)$ part of the deuteron operator, 
Tr$(G+G^\dagger)$, acquires a vacuum expectation value so that the operator can be replaced by $d_0 \sim 
\exp(\ri\sqrt{6\pi}\phi)$.  The resulting scaling dimension is $d_{deut} = 
3\tilde K/2,$ yielding a singular susceptibility as $T \rightarrow 0$ for $\tilde K \leq 2/3$.

The $2k_F$ meson density wave can also be found here.
Once the isospin sector is gapped $\la \mbox{Tr} G\ra \neq 0$, the scaling 
dimension of this order parameter is 
$
d_{DW} = 1/6\tilde K.
$
Hence at $1/6 <\tilde K < 2/3$ both the meson density wave and deuteronic superconductor susceptibilities are 
singular as $T\rightarrow 0$.

\section{ Conclusions}

 Our results demonstrate  that $1+1$D  
NJL model reproduces many properties expected for its $3+1$D prototype 
\cite{witten}.   We see, for example, that the masses of mesons and baryons are comparable to each 
other, as found in $3+1$D.
We stress that we have obtained these results for a realistic numbers of colors
, i.e. $N_c =3$. This was possible due to our use of the TCSA 
approach, a technique heretofore never applied to this particular problem. 
This powerful method enabled us 
to obtain masses of the multi-quark bound states including the six-quark deuteron,  
a notable achievement in the study of non-integrable strongly correlated systems. Although we 
have not explored this possibility, our theory can describe bound states of 
twelve quarks or more; they may emerge in the presence of the t'Hooft term with 
$\lambda <0$.  The spectrum that we have obtained numerical estimates for
include stable fermionic solitons with quantum numbers of baryons:  
U(1) charge $\pm 1$, SU(2) isospin $I=1/2$ 
(nucleons) and $I=3/2$ ($\Delta$-baryons). There are also stable excitations 
with quantum numbers of mesons (U(1) neutral particles with isospin 0 and 1 and 
Lorentz spin 0), and six-quark bound states (isospin I =1 dibaryons and isospin 
I=0 deuterons).
 
A part of our work is related to the dense baryon matter. We have 
obtained the phase diagram which contains such phases as a baryon metal,
analogous to non-Fermi liquid metallic states of one-dimensional condensed 
matter models, and various states with isospin gap and quasi-long range order, 
such as $2k_F$ meson density wave as well as a superconducting phase of condensed
deuterons.

\section*{Acknowledgements}
The authors are grateful to D. Gepner, L. Glazman, D. Kharzeev, G. Korchemsky, L. McLerran, G. Mussardo, R. 
Pisarski, and A. Zamolodchikov for discussions and interest in the work.  AMT and 
RMK were supported by the U.S. Department of Energy (DOE), Division of Materials 
Science, under Contract No. DE-AC02-98CH10886. TP was supported by a 
postdoctoral fellowship from the Hungarian Academy of Sciences (HAS), while TP 
and GT were also partially supported by the Momentum grant LP2012-50 of the HAS.
PL would like to thank CNRS (France) for financial support (PICS grant).
\appendix

\section{Non-abelian bosonization}

The foundation of this method is the fact that the Hamiltonian
density of free Dirac fermions with symmetry 
U(1)$\times$SU($N$)$\times$SU($M$) can be represented as a sum of a
Gaussian theory and two 
Wess-Zumino-Novikov-Witten conformal field theories (WZNW CFT) 
models of levels $k=M$ and $k=N$ respectively 
\cite{knizhnik,book,affleckspinchain}:
\bea
 &&\sum_{j=1}^M\sum_{\s =1}^N\ri(-R^\dagger_{j\s}\p_x R_{j\s} + L^\dagger_{j\s}\p_x L_{j\s}) 
= 
 {\rm WZNW} [SU(N)_M] + {\rm WZNW} [SU(M)_N] \nonumber\\
 &&+ \frac{1}{2}\left[ (\p_x\theta)^2 + (\p_x\phi)^2\right]. \label{nonAb}
\eea
As a consequence, one can write down  $n$-point
correlation functions of fermions  as a linear combination of  
products of $n$-point holomorphic  and antiholomorphic  conformal blocks of a 
U(1)
Gaussian theory and the corresponding WZNW models. 
Hence, in the general case, such a decomposition must be understood not as an 
operator identity, as was the case for abelian bosonization, but only as an 
identity for the conformal blocks 
\cite{affleckspinchain,bookqcd}.  Nevertheless, such identities will emerge at 
low energies in the theory given in Eqn. \ref{sigma}
of the main text. 

 The Hamiltonian density of the SU($M$)$_N$  WZNW model  is a 
sum of bilinears of its chiral Kac-Moody   SU($M$)$_N$ currents $J^A, \bar J^A$:
\be
{\rm WZNW} [SU(M)_N] = \frac{2\pi}{N + M} \left[:J^A J^A: + :\bar J^A \bar 
J^A:\right],
\label{WZNWham}
\ee
where $:A:$ stands for the normal ordering of $A$. Each symmetry sector is decoupled.
This remains true even when the interaction terms in Eqn. \ref{model} are added as they are given in terms
of the currents of individual symmetry sectors.
When $M=2$, the SU(2)$_N$ WZNW model Hamiltonian can 
be decomposed further and represented 
as a sum of a Gaussian U(1) model describing the Cartan subalgebra and a
critical model of Z$_N$ parafermions CFT \cite{para}. 
We will need this decomposition in the discussion that follows.

 It is reasonable to suggest that as soon as the interaction dynamically 
generates mass in the  SU(3) sector that certain operators with zero Lorentz spin may 
acquire vacuum expectation values. Here we will consider the dynamically 
generated quark mass $M_q$ 
as very large.  It then follows that all physical operators in the low-energy 
sector must be SU(3) singlets. 
The model of Eqn. \ref{sigma} represents a projection of the initial NJL 
Hamiltonian onto this color singlet space. To obtain operators acting in the low 
energy sector,  we will need to project (\ref{mesonop},\ref{H32},\ref{H12}) onto 
the ground state $|{\rm GS} \ra$ of the integrable model 
\be
 {\cal H}_{\rm color} = {\rm WZNW} [SU(3)_2] + gJ^A \bar J^A =
 \frac{2\pi}{5}[:J^AJ^A: + :\bar J^A\bar J^A:] + gJ^A\bar J^A .\label{A3}
\ee
We claim that after such reduction Eqns. (\ref{mesonop},\ref{H32},\ref{H12}) 
become operator identities. The proof is similar to the one given by Reshetikhin 
and Smirnov who  performed the reduction of the sine-Gordon model to the minimal 
model with an integrable perturbation \cite{ReshSm}.

To perform the projection to the color singlet state we need to consider first 
the case when the bare quark mass and the instanton term are zero: $m^* =0, 
\lambda =0$. In this case, the sigma model in Eqn. (\ref{A3}) is gapless and critical and has 
an extended conformal symmetry. 
The underlying conformal field theory is
\[
[U(1)\times SU(2)_3]_R\times[U(1)\times SU(2)_3]_L .
\]
This symmetry is reduced to U(1)$\times$SU(2) as soon as  $m^* \neq 0$. 

 Let us define the following fields in terms of $\theta$ and its dual $\phi$:
\be
\varphi = (\phi + \theta)/2, ~~ \bar\varphi = (\theta - \phi)/2.
\ee
At $K=1$ (this corresponds to $g_f =0$) 
these fields are chiral. 
For future purposes we need to know that the vertex operator
\bea
V_{n,m} = \exp[\ri \sqrt{2\pi/3}(n\varphi + m\bar\varphi)]
\eea
has conformal dimensions for general $K$:
\bea
&& h = \frac{1}{48}\left[(n+m)/\sqrt K + (n-m)\sqrt K\right]^2, \nonumber\\
&&  \bar h = \frac{1}{48}\left[(n+m)/\sqrt K - (n-m)\sqrt K\right]^2,
\eea
so that their conformal (Lorentz) spin is independent of $K$:
\bea
h - \bar h = (n^2-m^2)/12.
\eea
We will see that we will need exactly such vertex operators to construct the operators creating
the mesons and the baryons.

Conformal blocks of the primary fields of the SU(N)$_k$ WZNW model transform according to various representations of the SU(N) group. Their conformal dimensions are related to the quadratic Casimir operators of the corresponding representations \cite{dms}:
\be
h_{rep} = \frac{C_{rep}}{k+N} \label{dimrep}.
\ee
In particular, the conformal blocks of the SU(2)$_3$ theory transforming according to the spin $j$ representation will be 
denoted as
\be
{\cal F}^{(j)}_{h}, ~~ h_j = \frac{j(j+1)}{5} ,
\ee
with $j=0,1/2,1,3/2$. As far as the blocks for the SU(3)$_2$ CFT are concerned, we will need only one of them, namely the one which transforms according to the representation with a single box Young tableau. The corresponding Casimir for the SU(N) is  $C_{rep} = (N -1/N)/2$, for $N=3$ it becomes 4/3 and the conformal dimension is 4/15.   

To illustrate the non-abelian bosonization procedure let us consider, for instance, the meson operators. They are bilinear in the fermionic operators, made from the right- and left- moving quarks. Hence they are bosonic operators with Lorentz spin 0. The symmetry considerations suggest that a meson  operator must be a product of a bosonic exponent of field $\theta$ and primary fields from the SU(2) and SU(3) groups respectively. Moreover, these fields have to transform according to the representations corresponding to the single box Young tableau:
\be
R_{j \alpha}^+ L_{k \beta} = \re^{-\ri\sqrt{2\pi/3}\theta}G_{\alpha\beta}D_{jk},
\ee
where $G$ is the matrix of Eqn. (\ref{sigma}) which transforms in the $j=1/2$ 
representation of the SU(2) isospin group and $D$ is the SU(3) fundamental matrix field.
According to (\ref{dimrep}) the conformal dimensions of these matrix fields for SU(N)$_k$ model is given by the relation
\[
h = \frac{N^2-1}{2N(k+N)},
\]
so that for the $G$ matrix it is 3/20 and the SU(3) matrix field $D$ it is  4/15.  Then at 
$K=1$ (the noninteracting theory) the sum of all three dimensions is 1/2 as it must be. When the quarks becomes massive the operator from the SU(3) sector acquires a finite vacuum average. Thus for the mesons we have
\be
{\vec M} = \la {\rm GS} |R_{j \alpha}^+{\vec \sigma}_{\alpha \beta} L_{j \beta} 
|{\rm GS} \ra \sim \re^{-\ri\sqrt{2\pi/3}\theta}\mbox{Tr}[\vec\s \; G^+] .
\ee

Now let us consider the operator for the right-moving 
nucleon. This operator is made of three quark fields. Since nucleon is a fermion, it has Lorentz spin 1/2 and therefore 
must contain two right- and  one left-moving quarks.  The bare conformal dimensions of such an operator is (1,1/2). 
The nucleon has isospin 1/2, hence the primary field in the SU(2) sector must transform as $j=1$ in the holomorphic and as 
$j=1/2$ in the antiholomorphic sector. 
The conformal dimensions of such operator are (2/5,3/20). In the SU(3) sector 
we have a color singlet. 
Here we suggest it is described by conformal blocks of the SU(3) matrix field $D$ with conformal dimensions (4/15,4/15). The result is
\bea
\epsilon^{abc}
R_{a\alpha}R_{b\beta}L_{c\gamma} = \exp[\ri\sqrt{2\pi/3}(2\varphi - \bar\varphi)]
{\cal F}^{(1)}_{2/5}{\cal\bar F}^{(1/2)}_{3/20}\mbox{Tr}D .
\eea
It is easy to check that the conformal dimensions of the left hand side coincide with the ones of the right hand side. 
The projection on the color singlet state where Tr$D$ condenses yields
\be
 n_{1/2}^{\alpha\beta\gamma} = \epsilon^{abc}\la {\rm GS}|
R_{a\alpha}R_{b\beta}L_{c\gamma}|{\rm GS}\ra 
\sim  \exp[\ri\sqrt{2\pi/3}(2\varphi - \bar\varphi)]{\cal 
F}^{(1,1/2)}_{2/5,3/20} .
\label{rightnucleonfield}
\ee
We can repeat the same analysis for the $\Delta$-baryon operator with isospin $I=3/2$ of Eq. (\ref{H32}). 
Since the latter is made of three right-moving quarks in an SU(3) singlet state,  we find:
\be
\Delta_{3/2}^{\alpha\beta\gamma} = 
\epsilon^{abc}R_{a\alpha}R_{b\beta}R_{c\gamma} \sim 
\exp(\ri\sqrt{6\pi}\varphi){\cal F}^{(3/2)}_{3/4} I_{SU(3)},
\label{DeltabaryonApp}
\ee
where $I_{SU(3)}$ is the identity field of the SU(3)$_2$ CFT. By projecting out the color singlet state, 
the identification  (\ref{H32}) is then reproduced.

At this point, it is interesting to express the baryonic operators in terms of Z$_3$ parafermionic fields by 
exploiting the coset construction: SU(2)$_3$/U(1) $\sim$ Z$_3$ \cite{para}.
The SU(2)$_3$ primaries ($\Phi_{m_j,\bar m_j}^{j, \bar j}$) are related to the Z$_3$ parafermionic ones 
($f_{m, \bar m}^{l, \bar l} $, $ m = 2 m_j, \bar m = 2 \bar m_j, l = 2j, \bar l = 2 \bar j$)  by \cite{para}:
\begin{equation}
\Phi_{m_j,\bar m_j}^{j, \bar j} = f_{m, \bar m}^{l, \bar l} :\exp\left(i   m \sqrt{\frac{2\pi}{3}}\; \varphi_s
+ i   \bar m \sqrt{\frac{2\pi}{3}}\; \bar \varphi_s \right):,
\label{suparaprim}
\end{equation}
where $ \varphi_s$ and $\bar \varphi_s$ are the chiral components of a free bosonic field which accounts for the U(1) sector in the coset description. The Z$_3$  primary fields are subject to the constraints:
$f_{m,\bar m}^{l,\bar l} = f_{m +3,\bar m +3}^{3-l,3-\bar l}, ~~ f_{m,\bar m}^{l,\bar l} = f_{6+m, 6 + \bar m}^{l,\bar l}$,
where we take m to have periodicity 6, i.e. $m+6 \equiv m$.
The $\Delta$-baryon operator  (\ref{H32}) with isospin $I=3/2$ can then be expressed as a quartet field:
\bea
&& \Delta_{3/2,m} \sim  \exp \left( \ri\sqrt{6\pi} \; \varphi \right) F_m,  \label{3/2}\\
&& F_{m = \pm 3/2} = \exp \left( \pm\ri\sqrt{6\pi} \; \varphi_s\right), 
F_{m = \pm 1/2} = \exp  \left(  \pm\ri\sqrt{2\pi/3} \; \varphi_s\right) (\Psi, \Psi^+), 
\label{Deltabaryonpara}
\eea
where $\Psi$ is the Z$_3$ parafermionic current with holomorphic weight $2/3$.
By introducing the isospin projection from 
\be
I^z =  \sqrt{\frac{3}{2\pi}}\int \rd x \; \p_x \Phi_s  ,
\label{isospinprojbose}
\ee
$\Phi_s = \varphi_s + \bar\varphi_s$ being the total bosonic field,  one can check that 
Eq. (\ref{Deltabaryonpara}) has the correct isospin projection quantum numbers.

We now consider the right-moving nucleon field $n_{1/2}$ (\ref{rightnucleonfield}) with isospin $I=1/2$ 
and Lorentz spin $1/2$. By using the identification (\ref{suparaprim}), we find the following description for this doublet in 
the  Z$_3$  parafermionic basis:
\be 
n_{1/2, m=\pm 1/2} \sim  \exp[\ri\sqrt{2\pi/3}(2\varphi - \bar\varphi)]  \exp[\pm\ri\sqrt{2\pi/3}(2\varphi_s + 
\bar\varphi_s)](\mu,\mu^\dagger) ,
\label{nucleonparafermion}
\ee
where $\mu$ is the Z$_3$ disorder field with scaling dimension $2/15$ \cite{para}.
Using the bosonized expression of the isospin projection (\ref{isospinprojbose}), one can readily check that 
the operator (\ref{nucleonparafermion}) has the correct quantum numbers, i.e., isospin projection $m = \pm 1/2$ 
and Lorentz spin $1/2$.

\section{Quantum numbers of baryons}

We analyze in greater depth the excitations that interpolate between the different 
vacua of the semi-classical potential of Eq. (\ref{V}) of the main
text. There are two kinds of
solitons. Type 1 solitons are those which interpolate 
between the vacua corresponding to different signs of Tr$G$.  
These particles carry both U(1) charge and isospin and so are baryons. The 
corresponding operators are given in Eqs. (\ref{hadron}) and (\ref{3/2}).
Then there are type 2 solitons with isospin zero; they interpolate between
the degenerate minima of the bosonic exponent $(2\pi/3)^{1/2}\theta
\rightarrow (2\pi/3)^{1/2}\theta  + 2\pi n$.  The operator creating this
particle is given by Eq. (\ref{delta0}) of the main text.  Type 2
solitons have baryon number 2.

We define the baryon U(1) charge as one.  It is determined by the operator,
\be
\hat Q = \sqrt{\frac{2}{3\pi}}\int \rd x \p_x\theta .
\ee
Let $|q\ra$ be an eigenfunction of $\hat Q$ with eigenvalue $q$. Then using the 
identity 
\be
\hat Q \re^{\ri 2 n\sqrt{2\pi/3} \varphi(x)}|q\ra = (q + 2n/3)\re^{\ri 2n\sqrt{2\pi/3} \varphi(x)}|q\ra ,
\ee
we establish that the above bosonic exponent of the chiral field $\varphi$ raises the 
charge by $2n/3$. 
The baryon operator (\ref{hadron}) has $n = 3/2$ and hence creates a state of 
charge one.  Therefore we conclude that this operator has a matrix element 
between the vacuum and a type 1 soliton state. Thus such solitons in our model are coherent 
particles. 

Now we will show that the vacua Tr$G = \pm 2$ really have structure. This 
becomes manifest in the Z$_3$ parafermion representation. The mass term of Eqn. (\ref{sigma}) of the main text in this 
representation reads as follows using the identification (\ref{suparaprim}):
\be
\re^{\ri\sqrt{2\pi/3}\left( \theta + \Phi_s \right)} \sigma + H.c,
\ee
where $\sigma$ is the Z$_3$ spin field with scaling dimension $2/15$.

Transitions between minima of this effective potential give rise to solitons. 
Stable (anti)solitons  are charged ones  with U(1) charge $Q=\pm 1$. 
They  interpolate between the vacua as follows: 
\bea
 && \sqrt{2\pi/3}\theta \rightarrow \sqrt{2\pi/3}\theta \pm \pi, \\
 && \s^\dagger \rightarrow \re^{-2k\ri\pi/3}\s^\dagger, ~~ \s \rightarrow 
\re^{2k\ri\pi/3}\s,\nonumber\\
&& \sqrt{2\pi/3}\Phi_s \rightarrow \sqrt{2\pi/3}\Phi_s + \pi(\pm 1-2k/3), ~~ 
k=0,1,2 .\nonumber
\label{solitonstype1}
\eea

Using the identity (\ref{isospinprojbose}), these processes correspond to isospin projection 
$I^z = \pm 3/2 - k$.  From Eq. (\ref{solitonstype1}), we observe that the transformation on the $\sigma$ spin field corresponds 
to its Z$_3$ charge assignment \cite{para}.
Operators $F_{\pm 3/2}$ in Eq. (\ref{3/2}) 
correspond to $k=0$ and hence to $I^z = \pm 3/2$ as might be expected. The operator $F_{1/2}$
in Eq. (\ref{Deltabaryonpara}) has $k=1$ since the Z$_3$  parafermion current $\Psi$ carries a $k=1$ 
Z$_3$ charge \cite{para} and therefore corresponds to  $I^z = 1/2$ as it should.

As we have already said, there are also charge neutral particles corresponding 
to  
\bea
 && \sqrt{2\pi/3}\theta \rightarrow \sqrt{2\pi/3}\theta  \nonumber\\
 && \s^\dagger \rightarrow \re^{-2k\ri\pi/3}\s^\dagger, ~~ \s \rightarrow 
\re^{2k\ri\pi/3}\s,\nonumber\\
&& \sqrt{2\pi/3}\Phi_s \rightarrow \sqrt{2\pi/3}\Phi_s + 2\pi(1-k/3), ~~ 
k=0,1,2.
\eea 
In principle they can decay into soliton-antisoliton pairs of charged ones. 
Without numerical calculations their 
stability cannot be assured. Such particles, if they exist, are likely to be  
$I=1$ mesons (pions).
To look at dibaryons (charge 2) we can treat Tr$G$ in Eq. (\ref{V}) by a 
mean-field approximation, resulting in an effective double sine-Gordon model.

\section{Correlation functions}

\subsection{Zero density of baryons}
In the zero density phase the charge and the isospin sector are coupled and the 
baryons exist as coherent particles. The coherent parts of their Green's 
functions are fixed by the Lorentz symmetry.  For the nucleons we have the
following time-ordered Euclidean Green's function:
\begin{eqnarray}
\la T n_{a1/2}(x,\tau)n_{b1/2}^\dagger(0)\ra &=& \theta(\tau)\int \frac{d\ta}{2\pi}
e^{-m\tau \cosh(\ta) + imx\sinh(\theta)}e^{-\alpha_{1/2ab}\ta}\cr\cr
&& -\theta(-\tau)\int \frac{d\ta}{2\pi}
e^{m\tau \cosh(\ta) - imx\sinh(\theta)}e^{-\alpha_{1/2ab}\ta},
\end{eqnarray}
where the rapidity $\ta$ parameterize the energy, $m\cosh(\ta)$, and momentum, $m\sinh(\ta )$, of a
particle and $a,b=R,L$ indicates whether the nucleon is right or left moving.  
$e^{-\alpha_{1/2ab}\ta}$ is the matrix element squared between
a nucleon state with rapidity, $\ta$, and the nucleon field operator.  This matrix
element is determined by Lorentz symmetry to be:
$$
e^{\mp\frac{\ta}{2}} = \langle 0|n_{R/L,1/2}(0,0) |0\rangle,
$$
(up to a normalization constant).  The factor of $1/2$ in argument of the exponential reflects
the nucleons having Lorentz spin 1/2 and so $\alpha_{1/2ab}= \delta_{ab}$.
If we take the Fourier transform and then continue $\omega \rightarrow -i\omega + \epsilon$,
we obtain
\be
\la n_{a1/2}n_{b1/2}^\dagger\ra(\omega,q) = \frac{Z_{1/2}}{\omega^2 -\epsilon^2_{1/2}(q)}\Bigg(
\begin{array}{cc}
(q-\omega ) & -\frac{M_{1/2}\omega}{\epsilon_{1/2}(q)}\\
-\frac{M_{1/2}\omega}{\epsilon_{1/2}(q)} & -(\omega +q)
\end{array}
\Bigg)_{ab}, 
 ~~ a,b = R,L,
\ee
where $Z_{1/2}$ is an overall normalization factor, $M_{1/2}$ is the mass of the nucleon, and
$\epsilon_{1/2}(q) = \sqrt{M_{1/2}^2+q^2}$.

We can do the same for the $\Delta$-baryons.  Here the
relevant matrix element is 
$$
e^{\mp\frac{3\ta}{2}} = \langle 0|\Delta_{R,L,3/2}(0,0) |0\rangle.
$$
Fourier transforming as before we obtain 
\begin{eqnarray}
\la \Delta_{a3/2}\Delta_{b3/2}^\dagger\ra(\omega,q) &=&
\frac{Z_{3/2}}{2 M^2_{3/2}\epsilon_{3/2}(q)}\cr\cr
&&\hskip - 2.0in\times \Bigg(
\begin{array}{cc}
\frac{(\epsilon_{3/2}(q)-q)^3}{\epsilon_{3/2}(q)-\omega}-\frac{(\epsilon_{3/2}(q)+q)^3}{\epsilon_{3/2}(q)+\omega} & \frac{-2M_{3/2}^3\omega}{\omega^2-\epsilon^2_{3/2}(q)}\\
\frac{-2M_{3/2}^3\omega}{\omega^2-\epsilon^2_{3/2}(q)}&\frac{(\epsilon_{3/2}(q)+q)^3}{\epsilon_{3/2}(q)-\omega}-\frac{(\epsilon_{3/2}(q)-q)^3}{\epsilon_{3/2}(q)+\omega} 
\end{array}
\Bigg)_{ab}, 
 ~~ a,b = R,L.
\end{eqnarray}
We see that both the nucleon and $\Delta-$baryon Green's functions fall off as $1/\omega$.

\subsection{Finite density of baryons}
In the gapped phase of the dense baryon matter the spin 
excitations are gapped and the charge ones are not. The gapped excitations which 
determine the single baryon Green's functions are charge neutral solitons 
interpolating between different minima of the potential Tr$G$ = $\pm 2$. The corresponding
matrix elements (in the $SU(2)_3$ sector of the theory) between these excitations
are the baryon fields are again fixed by the Lorentz invariance:
\bea
&& \la \theta|{\cal F}^{(3/2)}_{3/4}(\tau,x) |0\ra \sim 
\re^{-3\theta/4}\re^{-M_{3/2}(\tau\cosh\theta + 
\ri(x/v_s)\sinh\theta)}\nonumber\\
&& \la \theta|{\cal F}^{(1)}_{2/5}{\cal F}^{(1/2)}_{3/20}(\tau,x) |0\ra \sim 
\re^{-\theta/4}\re^{-M_{1/2}(\tau\cosh\theta + \ri(x/v_s)\sinh\theta)} 
.\nonumber\\
&& \label{matrix}
\eea
Here we will consider only the part of the single baryon Green's function which 
originates from emission of this single massive soliton. 

Generally for nucleons and the $\Delta-$baryons the Green's functions become
(we focus only on the right-mover fields here)
\bea \label{ImG}
\la n_{R,1/2}(x,\tau) n^\dagger_{R,1/2}(0,0)\ra &=& 
  \frac{(\tau + \frac{ix}{v_s})^{1/4}}{(\tau - \frac{ix}{v_c})^{1/2}(\tau -
\frac{ix}{v_s})^{1/4}}\Big(\frac{\tau_0^2}{\tau^2 
+(\frac{x}{v_c})^2}\Big)^{\eta_{c,1/2}}K_{1/2}(M_{1/2}\rho) + ... ,\cr\cr
\la \Delta_{R,3/2}(x,\tau) \Delta^\dagger_{R,3/2}(0,0)\ra &=& 
  \frac{(\tau +  \frac{i x}{v_s})^{3/4}}{(\tau - \frac{i x}{v_c})^{3/2}(\tau - \frac{i x}{v_s})^{3/4}}
\Big(\frac{\tau_0^2}{\tau^2 +(\frac{x}{v_c})^2}\Big)^{\eta_{c,3/2}}K_{3/2}(M_{3/2}\rho) + ... ,\cr\cr
\rho^2 &=& \tau^2 +(\frac{x}{v_s})^2, 
\eea
where $\tau_0 \sim (k_Fv_c)^{-1}$ is the ultraviolet cut-off, the dots stand for 
contributions which include more than one massive particle (for instance, one 
soliton and mesons). 

We have been unable to perform the Fourier transform in full generality (i.e. $v_c\neq v_s$ and 
$\eta_{1/2,3/2}\neq 0$) and so we focus on the easiest case, $v_c=v_s=v, \eta_{1/2,3/2}=0$.
We note that while it is possible to consider the case of equal spin and charge velocities
when $\eta_{1/2,3/2}\neq 0$, this gives rise to unphysical behaviour in the response functions
at large $\omega$.  Considering then this simplest case (which would hold for very low densities
of baryons such that the velocities have not been renormalized) we find (with $v_c=v_s=1$):
\begin{eqnarray}
I_{1/2}(\omega,k) &\equiv& \int d\tau dx e^{i\omega\tau-ikx} \la n_{R,1/2}(x,\tau) n^\dagger_{R,1/2}(0,0)\ra;\cr\cr
I_{3/2}(\omega,k) &\equiv& \int d\tau dx e^{i\omega\tau-ikx} 
\la \Delta_{R,3/2}(x,\tau) \Delta^\dagger_{R,3/2}(0,0)\ra;\cr\cr
I_a(\omega, k) &=& -4\pi \bigg(\frac{k-i\omega}{k+i\omega}\bigg)^a 
\frac{(\omega^2+k^2)^a\Gamma(a+1)}{2^aM_a^{2+a}\Gamma(1+2a)}{_2}F_1(a+1,1,1+2a,-\frac{\omega^2+k^2}{M_a}),
\end{eqnarray}
where $a=1/2,3/2$ for the nucleons/$\Delta-$baryons respectively.

\begin{figure}[!ht]
\centering
\includegraphics[width=0.75\columnwidth,clip]{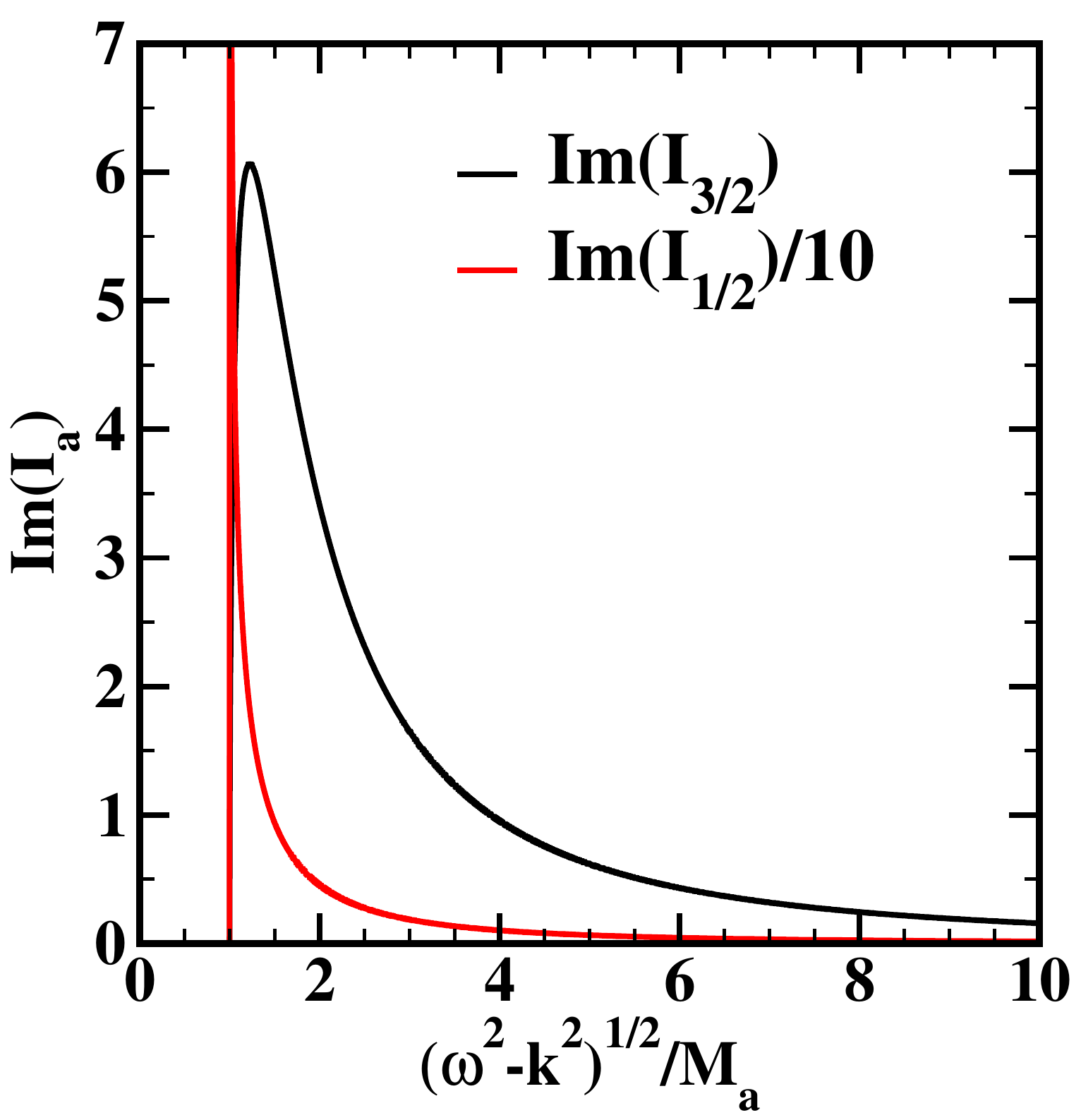}
\caption{The spectral weight of the baryon Green's functions plotted vs $\sqrt(\omega^2-k^2)/M_a$
 of the nucleons and $\Delta$-baryons.}
\label{fig:ImG}
\end{figure}

Continuing $\omega$ so as to obtain the retarded response, we plot the spectral weight (the imaginary
part of $I_a(-i\omega+\epsilon)$) for
each of the nucleon and $\Delta-$baryon in Fig. (\ref{fig:ImG}).  We see that this weight appears
as a broadened $\delta$-function.  In the zero baryon density phase, the spectral weight of the baryon
propagators are only seen at $\omega=k$.   But in the presence of finite baryon density, the baryons are
no longer coherent particles and this weight is smeared over a finite energy range.  We also the spectra function
of the nucleons is much more peaked than that of the $\Delta-$baryons, indicating the nucleons remain considerably
more coherent than their spin-3/2 counterparts in the presence of finite baryon density.

\end{document}